\documentclass[prb,preprint,floats,aps,superscriptaddress,showpacs]{revtex4-1}
\usepackage{amssymb}
\usepackage[squaren]{SIunits}
\usepackage{amsmath}
\usepackage{graphicx}
\usepackage{color}
\usepackage{multirow}
\usepackage{latexsym}

\usepackage{hyperref}

\def\k{{\mathbf k}}
\def\r{{\mathbf r}}

\def\r{{\mathbf r}}

\begin{document}
\date{\today}

%\graphicspath{{./FIGURE/}}
%%%%%%%%%%%%%%%%%%%%%%%%%%%%%%%%%%%%%%%%%%%%%%%%%%%%%%%%%%%%%%%%%%%
\title{Structure factors in granular experiments with homogeneous fluidization}

\author{Andrea Puglisi}
\affiliation{CNR-ISC UOS Roma Sapienza}
\affiliation{Dipartimento di Fisica, Universit\`a Sapienza - p.le A. Moro 2, 00185, Roma, Italy}

\author{Andrea Gnoli}
\affiliation{CNR-ISC UOS Roma Sapienza}
\affiliation{Dipartimento di Fisica, Universit\`a Sapienza - p.le A. Moro 2, 00185, Roma, Italy}
\affiliation{CNR-ISC UOS Roma Tor-Vergata}

\author{Giacomo Gradenigo}
\affiliation{CNR-ISC UOS Roma Sapienza}
\affiliation{Dipartimento di Fisica, Universit\`a Sapienza - p.le A. Moro 2, 00185, Roma, Italy}

\author{Alessandro Sarracino}
\affiliation{CNR-ISC UOS Roma Sapienza}
\affiliation{Dipartimento di Fisica, Universit\`a Sapienza - p.le A. Moro 2, 00185, Roma, Italy}

\author{Dario Villamaina}
\affiliation{Dipartimento di Fisica, Universit\`a Sapienza - p.le A. Moro 2, 00185, Roma, Italy}
\affiliation{CNR-ISC UOS Roma Sapienza}

\begin{abstract}
Velocity and density structure factors are measured over a
hydrodynamic range of scales in a horizontal quasi-2d fluidized
granular experiment, with packing fractions
$\phi\in[10\%,40\%]$. The fluidization is realized by
  vertically vibrating a rough plate, on top of which particles
perform a Brownian-like horizontal motion in addition to inelastic
collisions. On one hand, the density structure factor is
  equal to that of elastic hard spheres, except in the limit of large
  length-scales, as it occurs in the presence of an effective
  interaction. On the other hand, the velocity field shows a more
complex structure which is a genuine expression of a non-equilibrium
steady state and which can be compared to a recent fluctuating
hydrodynamic theory with non-equilibrium noise.  The temporal decay of
velocity modes autocorrelations is compatible with linear hydrodynamic
equations with rates dictated by viscous momentum diffusion,
corrected by a typical interaction time with the
thermostat. Equal-time velocity structure factors display a peculiar
shape with a plateau at large length-scales and another one at small
scales, marking two different temperatures: the ``bath'' temperature
$T_b$, depending on shaking parameters, and the ``granular''
temperature $T_g<T_b$, which is affected by collisions. The two ranges
of scales are separated by a correlation length which grows with
$\phi$, after proper rescaling with the mean free path.
\end{abstract}
% 45.70.-n = granular systems
% 05.40.-a = fluctuations
% 74.40.Gh = non-eq. processes
% 61.43.-j = Structure of disordered solids
\maketitle

%%%%%%%%%%%%%%%%%%%%%%%%%%%%%%%%%%%%%%%%%%%%%%%%%%%%%%%%%%%%%%%%%%%%%
\section{Introduction}
%%%%%%%%%%%%%%%%%%%%%%%%%%%%%%%%%%%%%%%%%%%%%%%%%%%%%%%%%%%%%%%%%%%%%

Understanding granular systems is strategically relevant
for the optimization of many industrial processes in pharmaceutical,
food, cosmetic, chemical, petroleum, polymer and ceramic
industries~\cite{intro1,JNB96b}. It is also essential for modeling
geophysical phenomena such as sediment fluidization in
landslides~\cite{intro_geo1,intro_geo2,intro_geo3} and soil
mechanics~\cite{intro_soil}. Finally, it is a challenge in the field
of statistical physics of out-of-equilibrium complex
materials~\cite{K99,kudrolli,irrevers,biroli,lengthscale,CD09}.
Experimental granular materials are usually limited to a
  few thousands particles, sometimes even less~\cite{G99,K99}. Because
  of this inherently ``small'' system size, fluctuations in granular
  experimental systems are rarely negligible and should be considered
  in the analysis.

Usually there is no general theory available to describe
  fluctuations in non-equilibrium stationary systems
  (NESS)~\cite{ML89,Z01,bsgj01}. A NESS is obtained in general by
  applying a steady driving force to a thermostatted system of
  interest: that results in a macroscopic breakdown of time-reversal
  accompanied by directed currents. The typical situation, within the
realm of molecular fluids, is by steady shearing -- giving place to a
transverse momentum current -- or by contact with two thermal
reservoirs at different temperatures -- leading to a steady current of
heat~\cite{SC85,OS06}. Therefore, molecular fluids are characterized
by non-equilibrium conditions associated in general to spatially
non-homogeneous configurations, i.e. currents appear in physical space
and break isotropy or other spatial symmetries.

An entirely different situation is present in so-called {\em active}
matter, which includes driven granular materials, as well as
``fluids'' of pedestrians, flocks of animals (birds or fish),
suspensions of bacteria or network of filaments (e.g. actin) in the
cell, etc.~\cite{TTR05,NRM07,JKPJ07,Cetal10} Common properties of
those systems are the presence of non-conservative interactions, often
mediated by a solvent, and of an external energy injection which can
be -- in principle -- spatially homogeneous (for instance, animal or
bacteria express their self-propulsion in the {\em bulk} of the system
and not only at its boundaries). The situation with granular materials
depends on the setup: experiments exist where most of the grains are
in contact with the external energy source irrespectively of their
position~\cite{OU98,PEU02,RIS07b,chile11}, resulting in a spatially
homogeneous NESS.

The present paper is devoted to investigate the structure of the
velocity field in an experimental setup, which realizes an homogeneous driving mechanism, already discussed
in~\cite{GSVP11b}. In homogeneous molecular fluids, the velocity field
is not expected to display any structure, as a consequence of the
separation of the Hamiltonian into a kinetic and a potential part
$\mathcal{H}=\mathcal{K}+\mathcal{U}$, the former being furtherly
separated in single-particle additive contributions,
$\mathcal{K}=\sum_i K_i$, that gives place to a factorized
Maxwell-Boltzmann kinetic distribution $\sim \prod_i \exp(-\beta
K_i)$, with no inter-particle velocity correlations and temperature
$1/\beta$. On the contrary, granular fluids are characterized by {\em
  non-conservative} interactions (kinetic energy is dissipated in
collisions), often resulting in inter-particle velocity
correlations, as well as velocity-position cross-correlations: such
effects are best appreciated in not too dilute setups, where they give
origin to visible non-equilibrium peculiarities, such as for instance
the violation of the Einstein fluctuation dissipation theorem
(FDT)~\cite{PBV07}.  Velocity correlations in a similar granular setup
have been previously studied in~\cite{PEU02}, measuring them in direct
space and -- therefore -- focusing on small spatial
distances. In this paper we shall focus on the study of
  structure factors (in $k$-space), since it highlights large
  space-time scales and is therefore more suitable to compare the
  experimental results with theory. Nevertheless the results
  of~\cite{PEU02} are compatible with those presented here (and in our
  previous publication~\cite{GSVP11b}), in the observation of
  exponentially space-dependent velocity correlations for the case of
  a rough vibrating plate.

Our experiments are also interesting to discriminate between two
models of ``granular thermostats'' which differ for the
presence/absence of a viscous drag force (per unit of mass) $-\gamma_b
{\bf v}_i$ proportional to the velocity ${\bf v}_i$ of the $i$-th particle: the presence of such a term allows one
to define a typical interaction time $1/\gamma_b$ with the thermostat
and a finite ``bath'' temperature $T_b$ (see the discussion in
Section~\ref{compare}). The experiment with a rough plate is well
described by a model with $\gamma_b>0$, and an effective bath
temperature can indeed be measured by looking at scales larger than a
well-defined correlation length.

This paper has the following organization: in Section~\ref{s:exp} the
experimental setup and the protocols of measurements are explained. In
Section~\ref{s:pdf} the probability distributions of velocities are
presented. The measurements of density, transverse and longitudinal
velocity structure factors are discussed in Section~\ref{sf}. From a
comparison of those structure factors with a phenomenological
fluctuating hydrodynamic theory, we get access to transport
coefficients, bath temperature and some typical non-equilibrium
correlation lengths. The dependence of those quantities with external
parameters are investigated in Section~\ref{temp}. Concluding remarks
appear in Section~\ref{s:concl}.

%%%%%%%%%%%%%%%%%%%%%%%%%%%%%%%%%%%%%%%%%%%%%%%%%%%%%%%%%%%%%%%%%%%%%%%%
\section{Experimental details}
\label{s:exp}

%-------------------------------------------------------------------
%                          fig 1
%-------------------------------------------------------------------
\begin{figure}[htbp]
\begin{center}
\includegraphics[width=6cm,keepaspectratio,clip=true]{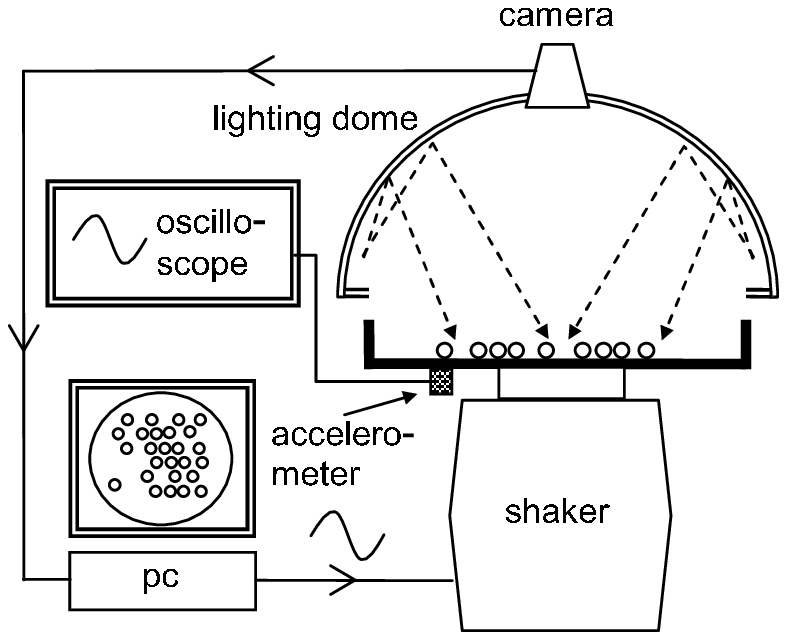}
\includegraphics[width=4cm,keepaspectratio,clip=true]{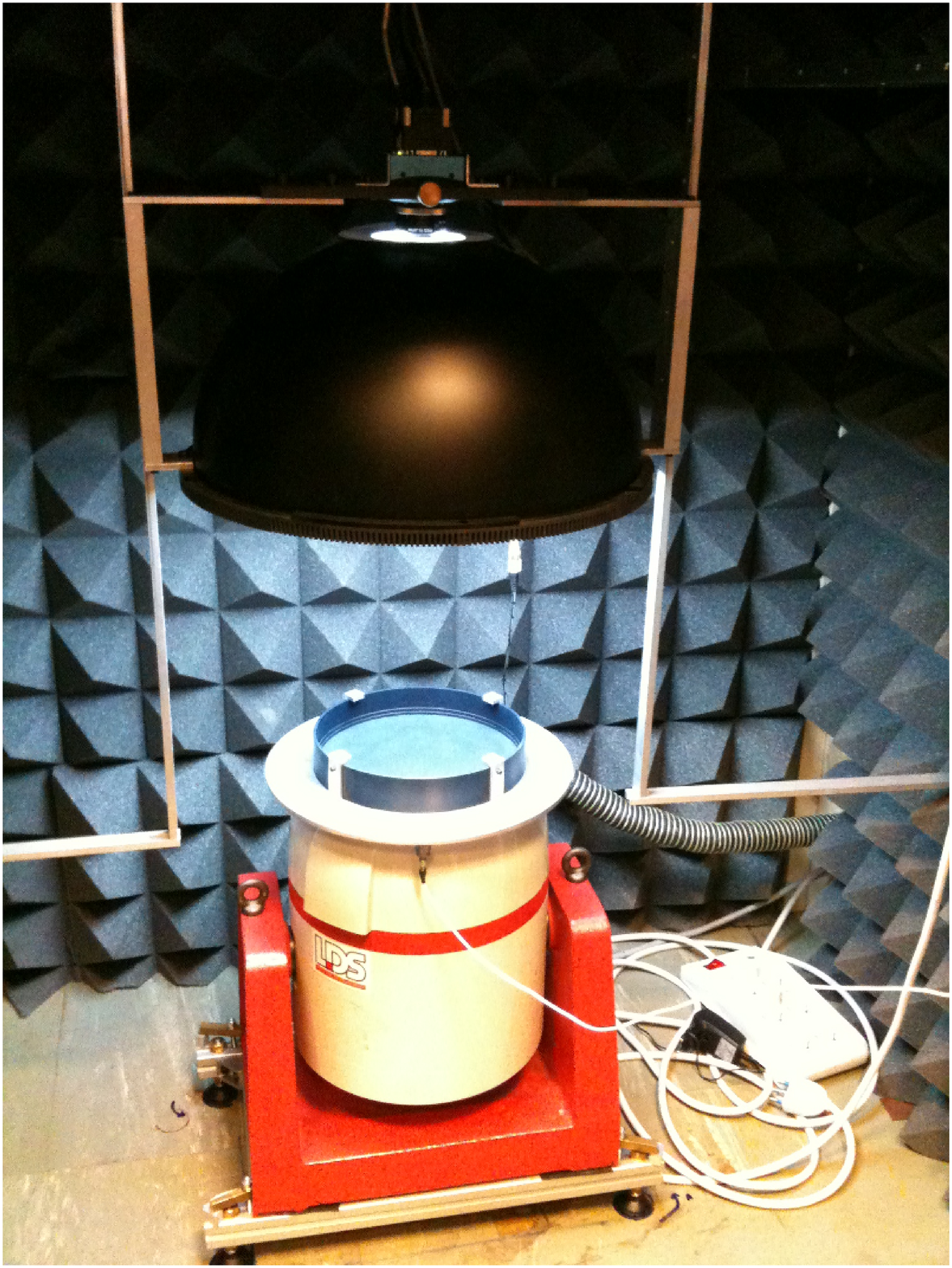}
\includegraphics[width=4cm,keepaspectratio,clip=true]{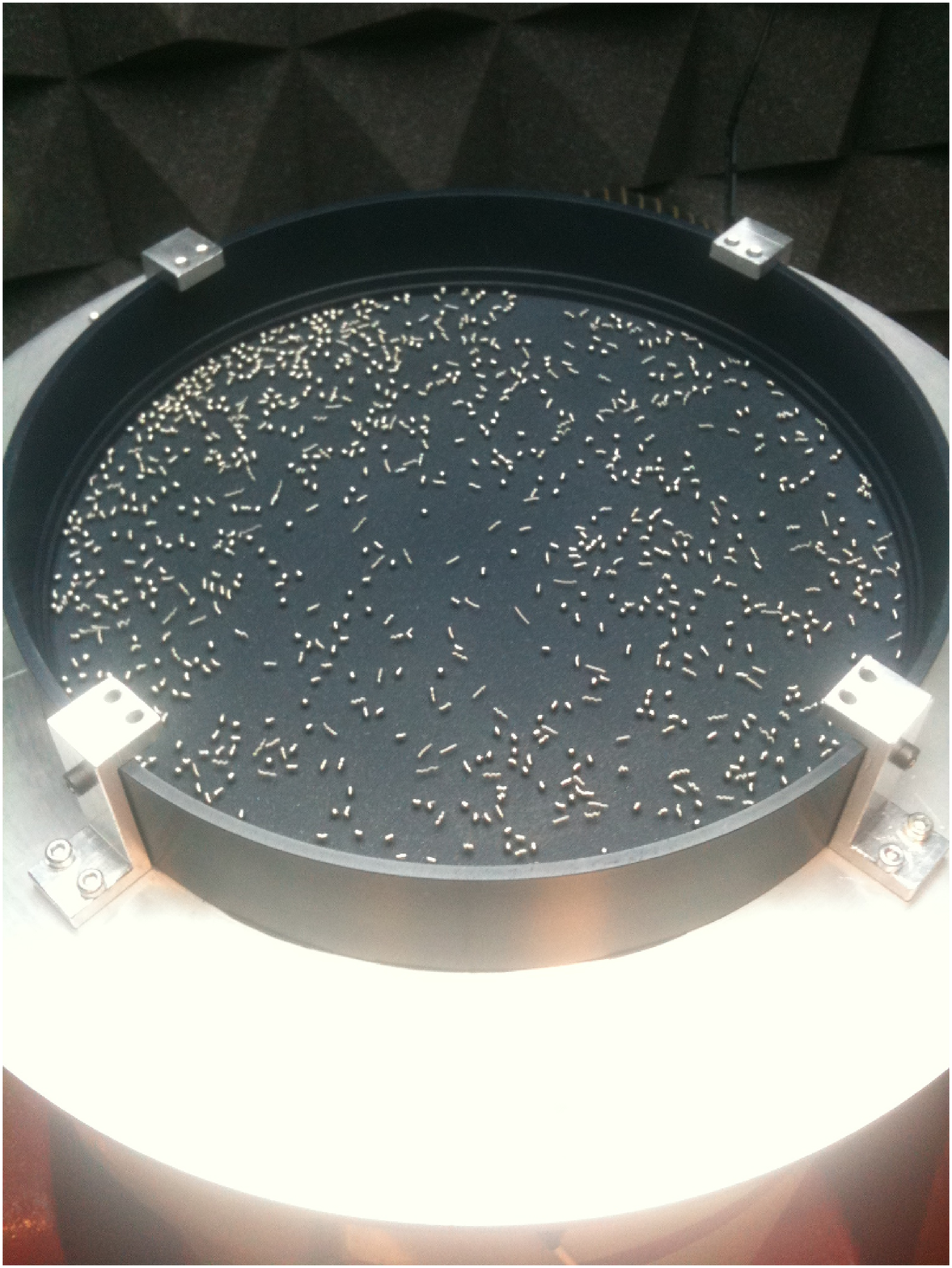}
\end{center}
\caption{Left: a logical scheme of the experiment. Center and right: two pictures of the experimental setup.}
\label{f:schema}
\end{figure}
%-------------------------------------------------------------------
%-------------------------------------------------------------------

In the left frame of Fig.~\ref{f:schema} a rationale of the experiment
is presented. The center and right frames reproduce real photographs
of the setup. See~\cite{video1} and~\cite{video2} for movies representing the time-evolution of
the system with different experimental parameters. The energy source is an
electrodynamic shaker, model V450 by LDS Test \& Measurement, placed
on top of a support with tunable inclination by means of adjustable
feet.  An aluminum mounting plate is placed over the head of the
shaker. The container of the granular media, which is fixed over the
mounting plate, is a circular aluminum plate with diameter
$D=\unit{200}{\milli\meter}$, with perpendicular borders (whose
height, \unit{20}{\milli\meter}, is however irrelevant for the present
setup).  The final horizontal alignment is verified, directly on top
of the plate carrying the granular sample, by means of a spirit level
with a precision of \unit{0.02}{\milli\meter\per\meter}. The alignment
check is repeated after every single run. The container has been
produced to have a flat floor and then it has been submitted to
sandblasting procedure with corundum mixture (aluminum oxide) whose
nominal grain size is \unit{1190}{\micro\meter}, also called ``gr16'':
actually, such a sand is highly polydisperse (grains as small as
\unit{200}{\micro\meter} can be found in it), and the rough surface
created in the sandblasting process presents {\em holes} with sizes in
the range [\unit{50}{\micro\meter}, \unit{500}{\micro\meter}], as
checked by means of an optical microscope. Finally, a black coating is
applied by electrochemical burnishing in order to enhance contrast for
particle visualization. The shaker is fed, through an amplifier PA1000
(LDS), with a sinusoidal signal at frequency $f$, which is converted
in a sinusoidal force applied to the payload. This appears as a
sinusoidal movement of the plate with amplitude $A$, i.e.
$z(t)=A*\sin(2\pi f t)$, where $z(t)$ is the vertical coordinate of
the plate at time $t$. Provided a linear response of the amplifier and
of the shaker, a given maximum amplitude of the input signal together
with a given mass of the payload determine a maximum acceleration
$\ddot{z}_{max}=A(2\pi f)^2$. For instance, with all other parameters
kept constant, a change of frequency results in a change of amplitude
of oscillation $A \sim 1/f^2$. Our experiments are done at frequencies
in the range \unit{[150,250]}{\hertz} and accelerations, in units of
gravity acceleration $g$, in the range $[5,11]$.   We have
  verified that, within those limits, the system remains in a quasi-2d
  regime, with a negligible number of particles flying at heights
  larger than their diameter.  Peak to peak acceleration is measured
during the course of each run (i.e. with full payload, including
moving particles) by means of an accelerometer, placed on the mounting
plate, connected to a digital averaging oscilloscope in order to
remove the spikes due to particle-plate collisions.

\begin{figure}[htbp]
\begin{center}
\includegraphics[width=8cm,keepaspectratio,clip=true]{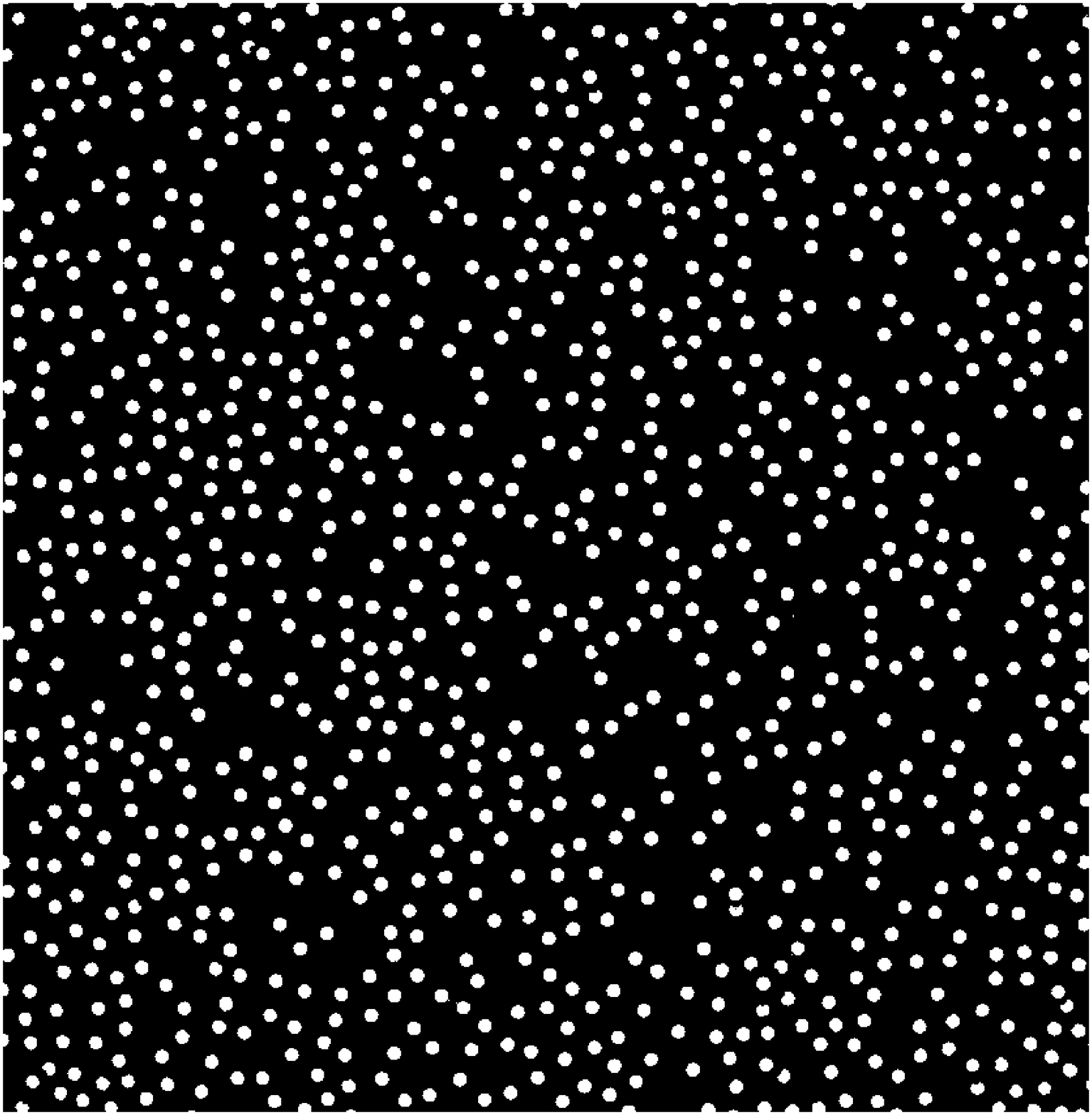}
\includegraphics[width=8cm,keepaspectratio,clip=true]{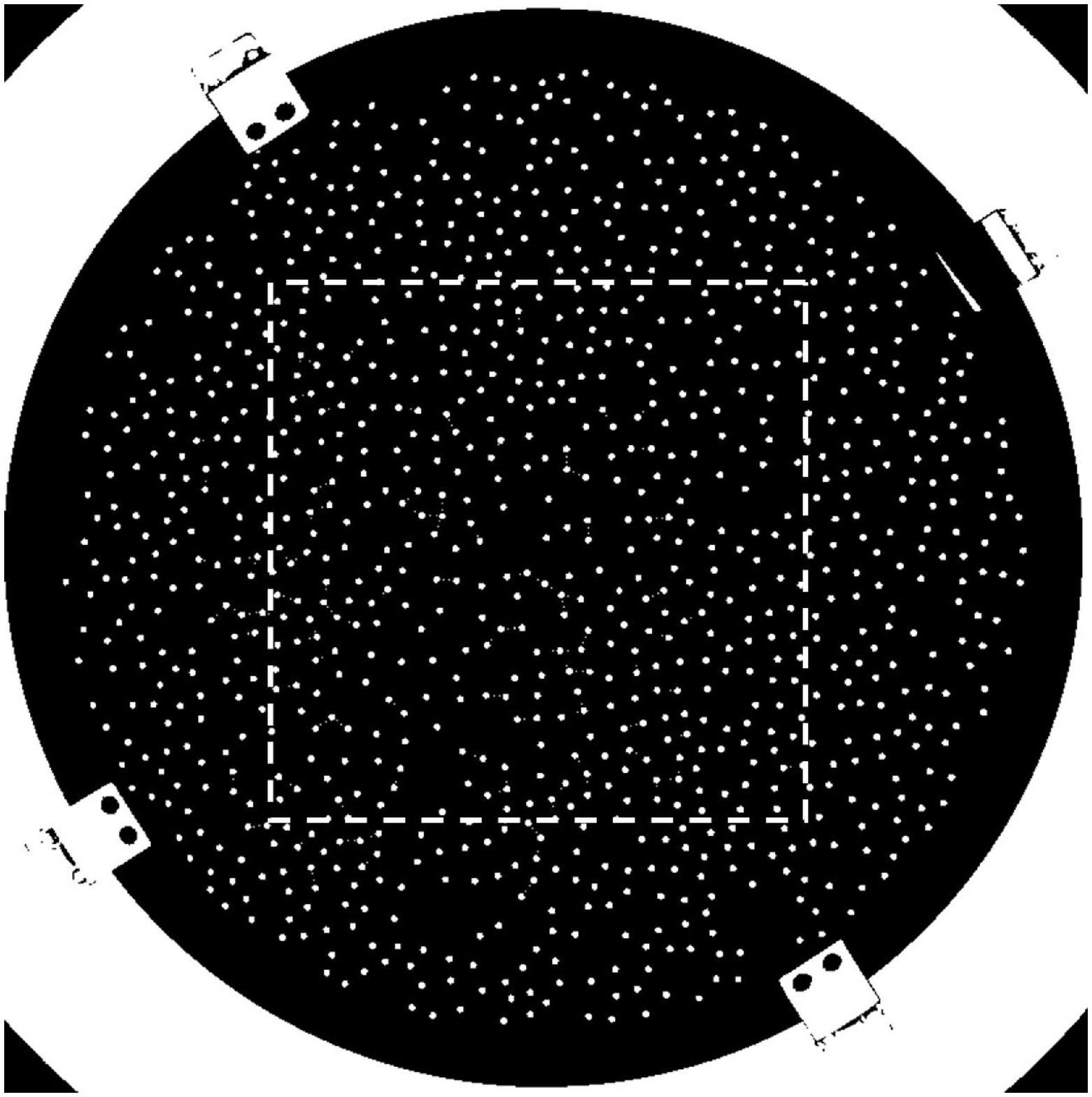}
\end{center}
% ho cambiato qualcosa nella caption
\caption{Two pictures from the acquisition camera (enhanced online). Left: zoom on the
  central region of interest. Right: whole system (refer also to
  Table~\ref{t:times}). The dashed box marks the zoomed region.}
\label{f:foto}
\end{figure}

The imaging apparatus consists of a fast camera, placed above the
central point of the container, and a lighting dome: the former is an
EoSens CL from Mikroron, the latter, placed at height much
  above the layer of particles (and therefore not interacting with
  them), is made of a led ring whose emitted light is reflected by a
diffusing half-spherical surface, reaching the granular medium with a
high degree of isotropy and uniformity.  As seen in the right frame of
Fig.~\ref{f:foto}, a very weak radial density gradient is present, so
that particles in the central region are slightly less concentrated
than those near the border. For the sake of verifying the importance
of such a weak spatial inhomogeneity, we have performed measurements
both in the whole container (e.g. right frame of Fig.~\ref{f:foto})
and in a restricted region focusing on a central rectangular ``region
of interest'' (roi), of size $L_x \times L_y =
\unit{100}{\milli\meter} \times \unit{100}{\milli\meter}$ (e.g. left
frame of Fig.~\ref{f:foto}). Changing the optics and the distance of
the camera, we manage to get, in both cases, a digital image with
dimensions $840\times 840$ pixels. Results obtained with the two
different ``roi'', shown in the following sections, do not display
qualitative differences but only weak quantitative discrepancies due
to small non-homogeneities. Nevertheless, a larger ``roi'' allows one
to probe modes with larger wavelengths, which is relevant in the
comparison with theoretical predictions for structure factors at small
wave-vectors. A very weak radial density gradient is visible in Fig.~\ref{f:foto}, due to the presence of hard boundaries where dissipation is stronger and density may be slightly enhanced.

Particles used in this experiment are restricted to a single material
and a few sizes, varying packing fraction, i.e. varying their number
$N$, and shaking parameters. Here, we use spheres with diameter
$\sigma$ in the range $[1,4]$ mm made of ``$316$-steel'', which is
non-magnetic in order to avoid interference with the shaker
electromagnet. Before each experimental run, particles and container are washed
in a ultrasound machine to eliminate those impurities which spoil the
reproducibility of the experiment: indeed, the effectiveness of energy transfer between
vertical vibration and horizontal spheres motion is mediated by
frictional properties of the sphere-surface contact, which could be
changed by cleanness of the apparatus.

Steel reflectance properties are such that each particle appears as
a bright and well defined spot, smaller than its diameter. A simple
lookup table is sufficient to segment the image distinguishing
the interesting white blobs (particles) from the black background. Particles
appear as almost-circular blobs with an average diameter $d_{blob}$ --
reported in Table~\ref{t:times} -- ranging from $\sim 2$ to $\sim 8$
pixels. The frame-rate $f_{acq}$ used for image acquisition is in between
$[60,250]$ frames per second. Reasons for the choice of space and time
resolution are discussed below.

%-------------------------------------------------------------------
%                          fig 2
%-------------------------------------------------------------------
%-------------------------------------------------------------------
%-------------------------------------------------------------------

As mentioned above,  the system of interest, here, is a
projection of the granular sample on the horizontal plane: only
horizontal positions and velocities are measured. It is directly
observed that particles never go on top of each other, allowing to
consider this system as a quasi-$2$-dimensional fluid, also in the
absence of a top lid.

The parameters used in the present paper are resumed in
Table~\ref{t:times}. To have an idea of the relevance of interactions,
we also report the mean free path $\lambda=1/(2\sqrt{\pi}g_2 n
\sigma)$, where $n$ is number density on the plane and $g_2$ the value
of the equilibrium pair correlation function at contact.
% qui ho cambiato un poco l'originale era: Each configuration is assigned a label in order to...
A label is assigned to each experimental configuration in order
to be easily referred to in the rest of the discussion. The rationale
of the set of experiment is the following: experiments A-H are done
with diameter \unit{2}{\milli\meter}, I-T are with diameter \unit{1}{\milli\meter}, U-X are with
diameter \unit{4}{\milli\meter}; experiments A-H and I-P focus on a central region,
while Q-X focus on the whole system; experiments A-D, I-L, Q-T and U-X
are groups of four configurations where only the packing fraction is
changed, from $10\%$ to $40\%$; experiments E-H and M-P are performed
at constant packing fraction, but varying the shaking parameters
(frequency or maximum acceleration).

\begin{table}
\begin{tabular}{|c||c|c|c|c||c|c|c|c||c|c|c|c|c|}
 \hline   Label  &$N$  & $\sigma$  & $\phi$  & $\lambda$  & $f$  & $z_{max}$  & $\dot{z}_{max}$  & $\frac{\ddot{z}_{max}}{g}$  & $f_{acq}$  & roi  & $d_{blob}$  & $\delta$  & $\delta$ \\[-6pt]
 & & (mm) & & (mm) & (Hz) & (mm) & (mm/s) &  & (Hz) & & (pixel) & (pixel) & (mm) \\
\hline
 \hline   A  &1000  &2  & 10\%  & 3.75  &  200  &0.039  & 49.1  &  6.3  &  250  & center  &8.368  &0.346  &0.041 \\ \hline
  B  &2000  &2  & 20\%  & 1.55  &  200  &0.039  & 49.1  &  6.3  &  250  & center  &8.368  &0.346  &0.041 \\ \hline
  C  &3000  &2  & 30\%  & 0.83  &  200  &0.039  & 49.5  &  6.3  &  250  & center  &8.368  &0.346  &0.041 \\ \hline
  D  &4000  &2  & 40\%  & 0.48  &  200  &0.039  & 49.5  &  6.3  &  250  & center  &8.368  &0.346  &0.041 \\ \hline
  E  &3000  &2  & 30\%  & 0.83  &  150  &0.070  & 65.5  &  6.3  &  250  & center  &8.368  &0.346  &0.041 \\ \hline
  F  &3000  &2  & 30\%  & 0.83  &  250  &0.026  & 40.6  &  6.5  &  250  & center  &8.368  &0.346  &0.041 \\ \hline
  G  &3000  &2  & 30\%  & 0.83  &  200  &0.048  & 60.8  &  7.8  &  250  & center  &8.368  &0.346  &0.041 \\ \hline
  H  &3000  &2  & 30\%  & 0.83  &  200  &0.031  & 39.0  &  5.0  &  250  & center  &8.368  &0.346  &0.041 \\ \hline
\hline  I  &4000  &1  & 10\%  & 1.88  &  200  &0.039  & 48.7  &  6.2  &  250  & center  &4.514  &0.471  &0.056 \\ \hline
  J  &8000  &1  & 20\%  & 0.78  &  200  &0.039  & 48.7  &  6.2  &  250  & center  &4.514  &0.471  &0.056 \\ \hline
  K  &12000  &1  & 30\%  & 0.42  &  200  &0.039  & 49.1  &  6.3  &  250  & center  &4.514  &0.471  &0.056 \\ \hline
  L  &16000  &1  & 40\%  & 0.24  &  200  &0.039  & 49.5  &  6.3  &  250  & center  &4.514  &0.471  &0.056 \\ \hline
  M  &12000  &1  & 30\%  & 0.42  &  150  &0.068  & 64.5  &  6.2  &  250  & center  &4.514  &0.471  &0.056 \\ \hline
  N  &12000  &1  & 30\%  & 0.42  &  250  &0.026  & 40.6  &  6.5  &  250  & center  &4.514  &0.471  &0.056 \\ \hline
  O  &12000  &1  & 30\%  & 0.42  &  200  &0.048  & 60.8  &  7.8  &  250  & center  &4.514  &0.471  &0.056 \\ \hline
  P  &12000  &1  & 30\%  & 0.42  &  200  &0.031  & 39.4  &  5.0  &  250  & center  &4.514  &0.471  &0.056 \\ \hline
\hline  Q  &4000  &1  & 10\%  & 1.88  &  200  &0.039  & 48.7  &  6.2  &  250  & whole  &2.257  &0.666  &0.164 \\ \hline
  R  &8000  &1  & 20\%  & 0.78  &  200  &0.039  & 49.1  &  6.3  &  250  & whole  &2.257  &0.666  &0.164 \\ \hline
  S  &12000  &1  & 30\%  & 0.42  &  200  &0.039  & 49.1  &  6.3  &  250  & whole  &2.257  &0.666  &0.164 \\ \hline
  T  &16000  &1  & 40\%  & 0.24  &  200  &0.039  & 49.1  &  6.3  &  250  & whole  &2.257  &0.666  &0.164 \\ \hline
\hline  U  &250  &4  & 10\%  & 7.51  &  200  &0.068  & 85.8  & 11.0  &   60  & whole  &7.979  &0.354  &0.087 \\ \hline
  V  &500  &4  & 20\%  & 3.11  &  200  &0.068  & 85.8  & 11.0  &   60  & whole  &7.979  &0.354  &0.087 \\ \hline
  W  &750  &4  & 30\%  & 1.67  &  200  &0.068  & 85.8  & 11.0  &   60  & whole  &7.979  &0.354  &0.087 \\ \hline
  X  &1000  &4  & 40\%  & 0.97  &  200  &0.068  & 85.8  & 11.0  &   60  & whole  &7.979  &0.354  &0.087 \\ \hline
\end{tabular}
\caption{Parameters of all the experiments discussed in the present
  paper. \label{t:times}}
\end{table}

As a matter of fact, for all choices of parameters considered here
particles perform an apparently random, Brownian-like motion on the
plane, with low flights interrupted by bounces on the plate or with
other particles. The study of a single particle on the plate,
unfortunately, gives information quite different from the case of a
many particle situation (e.g. at packing fraction $\ge 10\%$), which
is our system of interest. If particle's flights are not interrupted
by collisions with other particles, repeated bounces with the plate
may build up very high velocities (or resonances, in the
  case of a smooth surface)~\cite{LM93}: such a more general
situation illustrates the fact that the particle-plate friction
properties are influenced by the velocity of the particle itself
(e.g. multiplicative noise). Nevertheless, as put in evidence by the
comparison of our experiment with a model based on a
non-multiplicative noise thermostat, in the many-particles
configuration there is not the possibility of reaching such high
velocities and this velocity-dependent noise regime does not take
place.

%%%%%%%%%%%%%%%
\subsection{Acquisition of positions and uncertainty}

From each blob of pixels, the $x$ position of the center of mass of the particle
is obtained by averaging the average position of each row of pixels
(and the same is done for columns, obtaining the $y$
position). Uncertainty in the evaluation of the center of mass position is
estimated by a Monte-Carlo procedure where discs of diameter
$d_{blob}$ are placed on a grid of unitary pixels, choosing randomly
on a much finer scale (double precision, i.e. $10^{-8}$) the
coordinate of the disc center, and then their position is computed
using only covered pixels, with the procedure discussed before,
obtaining a distribution of errors. It is seen, as expected, that the
standard deviation of such a distribution is $\delta \sim
1/\sqrt{d_{blob}}$ which ranges -- as reported in Table~\ref{t:times} --
in $\sim [0.04-0.16] \unit{ }\milli\meter$, depending on the particular setup (the largest
error is obviously associated to the experiment done by observing the
full system with the smallest particles). It is immediately understood
that even increasing the linear resolution by a factor $2$ would not
reduce significantly this error.

%%%%%%%%%%%%%%%
\subsection{Assessment of ballistic and diffusive timescales}

\begin{figure}[htbp]
\begin{center}
\includegraphics[width=12cm,keepaspectratio,clip=true]{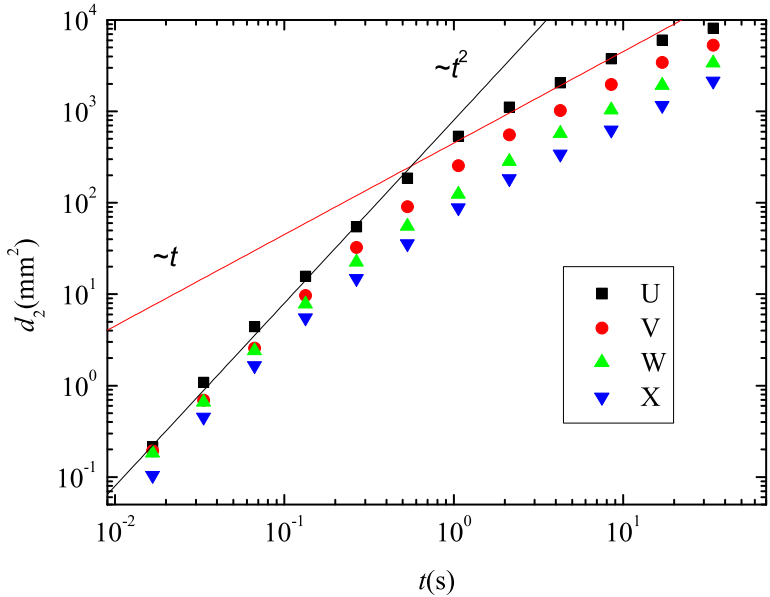}
\end{center}
\caption{Mean squared displacement $d_2(t)$ for different packing
  fractions $\phi=10\% (\blacksquare),20\% (\bullet),30\%
  ({\color{green}\blacktriangle}),40\%
  ({\color{blue}\blacktriangledown})$, with particles' radius
  $\sigma=4$ mm and for driving with $f=200$ Hz (experiments with
  labels U,V,W,X). For all other parameters, see
  Table~\ref{t:times}. Error bars fall within the
    symbols.}
\label{f:diff}
\end{figure}

At fixed spatial resolution, an {\em upper} limit in the choice of the
framerate $f_{acq}$ for image acquisition is immediately established,
given that our aim is computing {\em velocities} of particles. Indeed,
for a population of particles with typical velocity $v_0\sim
\sqrt{\langle v^2\rangle}$, it is pointless measuring displacements
between frames separated by a time interval $\Delta t = 1/f_{acq} <
\Delta t_{min}\equiv\delta/v_0$, because such a displacement would be
spoiled by the acquisition uncertainty $\delta$ previously
discussed. However, too large $\Delta t$'s would be wrong as well:
each particle moves ballistically until its flight is interrupted by
bounces with the plate or with other particles, therefore at large
$\Delta t > \Delta t_{diff}$ the displacement is diffusive and it does
not represent the real ``instantaneous'' velocity anymore.  For the
purpose of estimating the correct time-scale $\Delta t_{diff}$, we
first acquire data at high frame rate ($250$ frames
per second) and then compute the mean squared displacement as a
function of time interval $d_2(t)=\langle [x(t)-x(0)]^2 \rangle$:
ballistic ($d_2 \sim t^2$) and diffusive ($d_2 \sim t$) regimes appear
clearly, marking the time-scale $\Delta t_{diff}$ which separates the
two, see Fig.~\ref{f:diff}. We choose our final frame rate $f_{acq}$
for the experiment such that $\Delta t_{min} \ll 1/f_{acq} \ll \Delta
t_{diff}$ .

%-------------------------------------------------------------------
%                          fig 3
%-------------------------------------------------------------------
%-------------------------------------------------------------------
%-------------------------------------------------------------------

It is interesting to mention that the limit of the ballistic scale is
mostly independent of the period of oscillation of the plate, even
if it may be contrary to intuition. Our main interpretation of the
motion of particles on the rough vibrating surface, discussed in
Section~\ref{compare}, is that of a Brownian motion with viscous drag
determining a characteristic relaxation time $\tau_b=1/\gamma_b$, basically
related to the frictional properties of the sphere-surface contact,
to which superimpose inelastic collisions with other particles occurring
with a mean free time $\tau_c$. Such a model predicts a ballistic
regime at times smaller than $\min[\tau_c,\tau_b]$: quantitative
assessment of $\tau_b$ and $\tau_c$ are well consistent with the
estimate of $\Delta t_{diff}$ obtained here from the mean squared
displacement (see discussion in Section~\ref{temp}).

%%%%%%%%%%%%%%%%%%%%%%%%%%%%%%%%%%%%%%%%%%%%%%%%%%%%%%%%%%%%%%%%%%%%%%%%
%%%%%%%%%%%%%%%%%%%%%%%%%%%%%%%%%%%%%%%%%%%%%%%%%%%%%%%%%%%%%%%%%%%%%%%%
\section{Distributions of velocities}
\label{s:pdf}

After a tiny transient regime, our system reaches a stationary
homogeneous state, which is the subject of our research.  Equilibrium
statistical mechanics predicts a Gaussian (Maxwell-Boltzmann)
distribution for velocities in a fluid. The deviation of velocities in
granular fluids from the Maxwell-Boltzmann statistics is widely
discussed in the literature~\cite{NE98,PLMPV98,LCDKG99}. In the
presence of uniform driving, however, those deviations are expected to
be small and well reproduced by polynomial corrections superimposed to
the usual Gaussian statistics.

%-------------------------------------------------------------------
%                          fig 4
%-------------------------------------------------------------------
\begin{figure}[htbp]
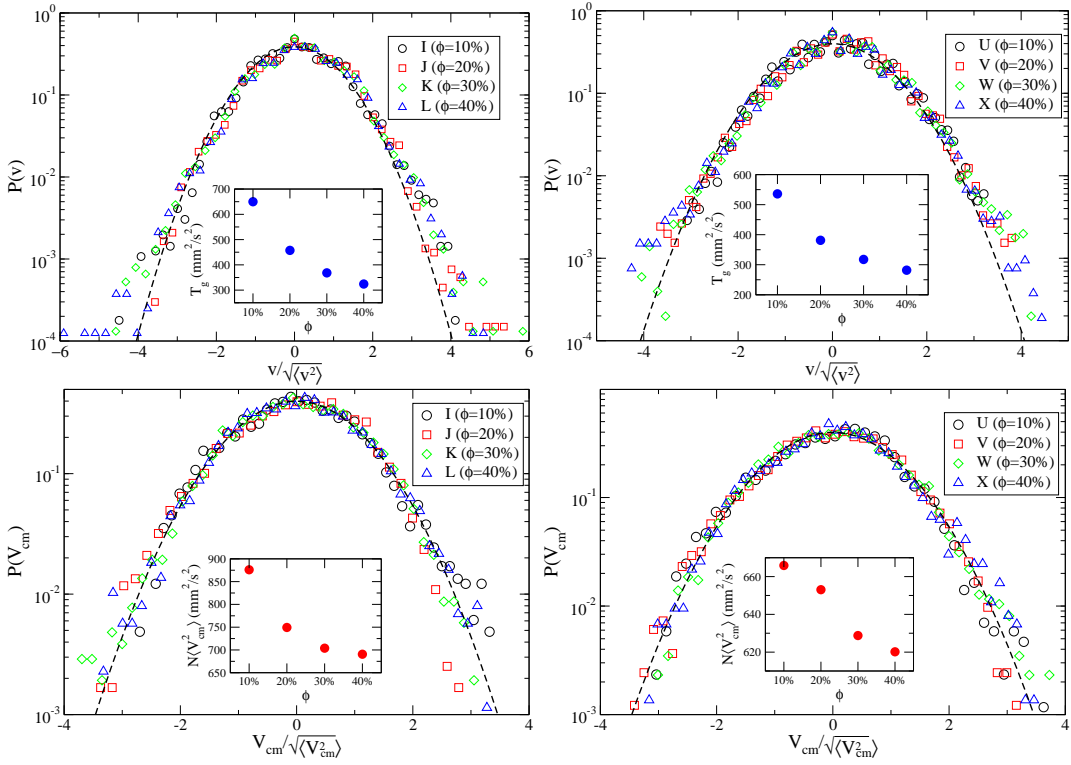

\begin{center}
\includegraphics[width=7cm,keepaspectratio,clip=true]{IJKL_pdf.eps}
\includegraphics[width=7cm,keepaspectratio,clip=true]{UVWX_pdf.eps}\\
\includegraphics[width=7cm,keepaspectratio,clip=true]{IJKL_pdf_cm.eps}
\includegraphics[width=7cm,keepaspectratio,clip=true]{UVWX_pdf_cm.eps}
\end{center}
\caption{Top: particle velocity distributions (rescaled to have unit
  variance) for experiments with different packing fractions $\phi$
  and particles of diameter $\sigma=1$ mm (left panels) and $\sigma=4$
  mm (right panels), for driving with $f=200$ Hz. Bottom: center of
  mass velocity distributions for the same experiments, again rescaled
  to have unit variance. Dashed black lines are normalized
  Gaussian. For all other parameters, see Table~\ref{t:times}. Insets:
  variances of the distributions, i.e. $T_g$ and $N\langle V_{cm}^2
  \rangle$, as functions of $\phi$. }
\label{f:pdf}
\end{figure}
%-------------------------------------------------------------------
%-------------------------------------------------------------------

In Fig.~\ref{f:pdf}, two top frames, experimental rescaled velocity distributions
are shown. Small but appreciable deviations from
the Gaussian ideal curve are visible, mainly in the
tails. Sonine-polynomial fits, not shown here, give a value for $a_2$
in the range $[0.01,0.05]$. The granular temperature $T_g=(1/2N)\sum_i
{\bf v}_i^2$ is also shown in the insets: it decreases with the
packing fraction $\phi$ as expected due to the increasing
collision rate.

Notice that, as stated in the formula above, in the following we measure
``temperatures'' in units of squared velocities, i.e. ignoring the mass
of the particle. Such a choice is possible because, here, temperature
has only a kinetic meaning and does not correspond to environmental
temperature (which occurs on energy scales exceedingly smaller).

%%%%%%%%%%%
\subsection{Distribution of the center of mass velocity}

In the two bottom frames of Fig.~\ref{f:pdf}, one can see the
distribution of one component ($x$ or $y$) of the center of mass
velocity ${\bf V}_{cm}(t)=(1/N)\sum_{i=1}^N {\bf v}_i(t)$, collected
over different configurations ($5\times 10^3$) in each experiment. It
is interesting to notice how the distribution is 
  well-fitted to a Gaussian. The rescaled variance $N\langle
V^2_{cm}\rangle$ (where by isotropy we define $V_{cm}^2={\bf
  V}_{cm}^2/2$) is also plotted in the inset, as a function of the
packing fraction. Notice that such a rescaled variance changes with
the packing fraction on a much smaller energy interval with respect
to the variation observed for $T_g$.

The effect of inter-particle collisions upon the center of mass
velocity is much smaller when compared with the velocity of the single
particle: indeed collisions conserve momentum for each component, so
that the total momentum in a region is only affected by bounces on the
plate, and by exchanges at the boundaries of the region as well
(e.g. particles going in or out from the region itself). The latter
contribution is as smaller as larger is the size of the region.

%%%%%%%%%%%%%%%%%%%%%%%%%%%%%%%%%%%%%%%%%%%%%%%%%%%%%%%%%%%%%%%%%%%%%%%%
%%%%%%%%%%%%%%%%%%%%%%%%%%%%%%%%%%%%%%%%%%%%%%%%%%%%%%%%%%%%%%%%%%%%%%%%
\section{Structure factors} 
\label{sf}

The main goal of our study is concerned with the characterization and
the analysis of large-scale fluctuations.  The physical quantities of
interest are the correlations between the coarse-grained hydrodynamic
fields, density $n({\bf r},t)$, velocity ${\bf u}({\bf r},t)$ and
temperature $T({\bf r},t)$, defined as follows:
\begin{eqnarray}
n({\bf r},t)&=&\sum_i\delta({\bf r}-{\bf r}_i(t)), \nonumber \\
{\bf u}({\bf r},t)&=&\frac{1}{n}\sum_i {\bf v}_i(t)\delta({\bf r}-{\bf r}_i(t)), \label{sf.0} \\
T({\bf r},t)&=&\frac{2m}{dn}\sum_i \frac{v^2_i(t)}{2}\delta({\bf r}-{\bf r}_i(t)), \nonumber 
\end{eqnarray}
where ${\bf r}_i(t)$ is the position of particle $i$ at time $t$, and the sum runs over the $N$ particles.
In particular, we focus on large spatial scale fluctuations of
the system around the stationary \emph{homogeneous} state, where the
hydrodynamic fields take the values $n=\overline{n}$, $T=\overline{T}$
and ${\bf u}=0$.  We study fluctuations in momentum space and define
$\delta {\bf a}({\bf k},t)=\{\delta n({\bf k},t),\delta T({\bf
  k},t),u_\parallel({\bf k},t),u_\perp({\bf k},t)\}$, with $\delta
{\bf a}={\bf a}-{\bf \overline{a}}$, where the Fourier transform is
\begin{equation}
\delta {\bf a}({\bf k},t)=\int d{\bf r}~\delta {\bf a}({\bf r},t)e^{- i{\bf k}\cdot{\bf r}},
\label{sf.1}
\end{equation} 
and $u_\perp({\bf k},t)$ and $u_\parallel({\bf k},t)$ denote the shear
and longitudinal modes, respectively, namely
\begin{eqnarray}
u_\perp({\bf k}) &=& \hat{k}_\perp \cdot {\bf u}({\bf k}) \nonumber \\
u_\parallel({\bf k}) &=& \hat{k}\cdot {\bf u}({\bf k}), 
\label{sf.2}
\end{eqnarray}
$\hat{k}_\perp$ being a unitary vector such that $\hat{k}_\perp \cdot
\hat{k}=0$.  In the following, we show and discuss the experimental
data obtained for the static and dynamic correlation functions of the
hydrodynamic fields, defined, in the stationary state, as
\begin{equation}
S_{ab}({\bf k})=\frac{1}{N}\langle \delta
a({\bf k})\delta b({-\bf k})\rangle,
\label{sf.3}
\end{equation}
and 
%% \begin{equation}
%% S_{ab}({\bf k},\omega)=\int_{-\infty}^{\infty}S_{ab}({\bf k},t)e^{-i\omega t},
%% \label{sf.4}
%% \end{equation}
%% with
\begin{equation}
S_{ab}({\bf k},t)= \frac{1}{N}
\langle \delta a({\bf k},t)\delta b({-\bf k},0)\rangle.
\label{sf.5}
\end{equation}

The study of these quantities provides important information on the
structural properties of the system and on the characteristic lengths
and times governing macroscopic fluctuations. In particular, at equilibrium and at small
$k$, the static structure factors, Eq.~(\ref{sf.3}), are related to
thermodynamic quantities~\cite{HMD96}.  From the measure of dynamic
structure factors, Eq.~(\ref{sf.5}), one can extract information on
heat and sound hydrodynamic poles, and thus on the transport
coefficients which rule thermal diffusion and propagation of
pressure waves in the system.

From the procedural point of view, we measure structure factors
$S_{ab}({\bf k},t)$ by directly applying
definitions~\eqref{sf.1}-\eqref{sf.5} on a set of wave-vectors ${\bf
  k}$, averaging over several (typically $5\times 10^3$)
configurations; subsequently, we compute sphericized averages
$S_{ab}(k,t)$ over shells with average modulus $k$, assuming
isotropy. The smallest value of $k$ considered is
  $k_{min}=2\pi/(D/2)$ accordingly to the largest linear size of the
  system (we recall that $D$ is the diameter of the container). In the
  case where we focus on the central region of width $L_x=D/2$, the
  minimum value of wave-vector is $k_{min}=2\pi/(L_x/2)$. For the
  largest value of $k$ we have chosen a value, typically smaller than
  the diameter of the particles, convenient with respect to the
  relevant information we want to highlight (e.g. for the velocity
  structure factor it is useful to give a convincing evidence of the
  reached ``plateau'' at large $k$).

%%%%%%%%%%%%%%%%%%%%%%%%%%%%%%%%%%%%%%
\subsection{Density structure factor}

From Eq.~(\ref{sf.3}) we get the density structure factor

\begin{equation}
S_{nn}({\bf k})=\frac{1}{N}\langle \delta n({\bf k})\delta n({-\bf k})\rangle 
=\frac{1}{N} \left\langle \sum_{i,j=1}^{N} \exp(-i \k \cdot (\r_i-\r_j)) \right\rangle,
%\\&=&\frac{1}{N} \sum_{i=1}^{N} \left\langle | \exp(-i \k \cdot \r_i) |^2 \right\rangle,
\label{sf.6}
\end{equation}
where the average is taken over the stationary state.

%-------------------------------------------------------------------
%                          fig 5
%-------------------------------------------------------------------
\begin{figure}[htbp]
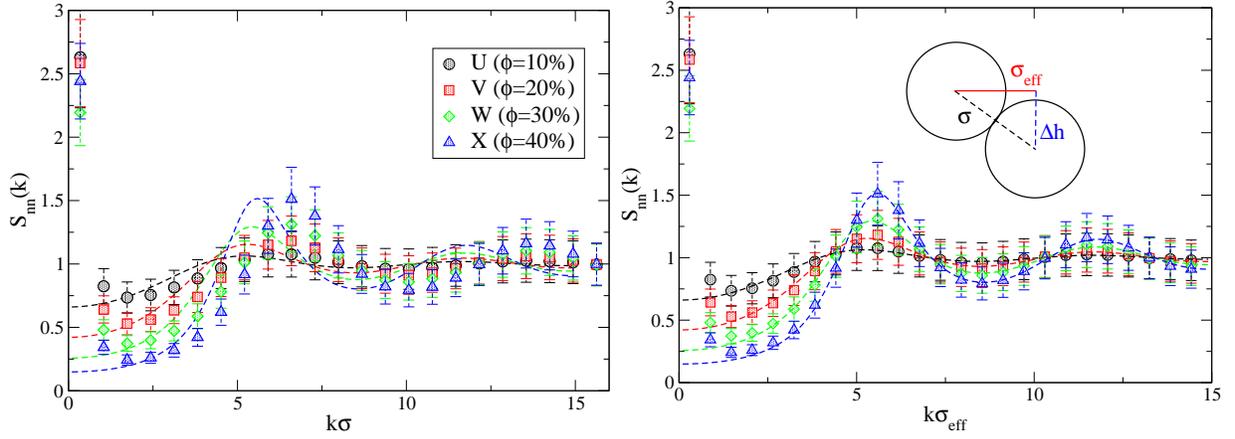

\begin{center}
\includegraphics[width=8cm,keepaspectratio,clip=true]{s_nn.eps}
\includegraphics[width=8cm,keepaspectratio,clip=true]{s_nn_eff.eps}
\end{center}
\caption{Density structure factors at various packing fractions $\phi$
  for particles with diameter $\sigma=4$ mm and for driving with
  $f=200$ Hz (experiments U,V,W,X). The dashed lines represent the
  analytical prediction for elastic hard disks~\cite{BC96}. In the
  graph on the left the experimental results are plotted versus
  $k\sigma$, while in the graph on the right they are shown versus
  $k\sigma_{eff}$ with $\sigma_{eff}/\sigma=cos(\pi/6)$. The cartoon
  in the second graph illustrates the meaning of $\sigma_{eff}$ which
  is the effective diameter seen by particles which collide at an
  average height $\approx \sigma/2$. For all experimental parameters,
  see Table~\ref{t:times}.}
\label{f:snn}
\end{figure}
%-------------------------------------------------------------------
%-------------------------------------------------------------------

In Fig.~\ref{f:snn} we show $S_{nn}(k)$ measured in our system,
together with the analytical prediction for an elastic system of hard
discs.  The comparison with the elastic system shows two main
differences. First, at small $k$ the structure factors displays an
enhancement, signalling long range density correlations
due to inelasticity. Those correlations are not the main objective of
our analysis here and have been addressed before, see for
instance~\cite{OU98,CDMP04,mex09}. Notice that such correlations at large
wavelengths do not depend monotonously on $\phi$: higher packing
fractions, as a matter of fact, enhance both the aggregating role of dissipation (due
to inelastic collisions) and the dispersive role of excluded volume. Therefore, they may
have negative or positive influence on the density structure. Second, at larger $k$ the
theoretical curves for elastic hard discs are in excellent agreement
with our experiment, when the abscissa is rescaled by a factor
$\cos(\pi/6)$, as shown in the right frame of Fig.~\ref{f:snn}. Such a
rescaling is equivalent to consider an {\em effective} diameter of
particles $\sigma_{eff} \approx \cos(\pi/6) \sigma$. Indeed, particles
typically collide at different heights from the plate and therefore
meet at a distance smaller than their diameter. Such an observation is
consistent with an average difference of height $\Delta h \approx
\sigma/2$.

%%%%%%%%%%%%%%%%%%%%%%%%%%%%%%%%%%%%%%%%%%%
\subsection{Transverse velocity modes}

As discussed below, in the realm of linear hydrodynamics, transverse
velocity modes are decoupled from the other hydrodynamic fields and
their behaviour can be studied separately.  From Eq.~(\ref{sf.3}) one
has
\begin{equation}
S_{\perp}({\bf k})=\frac{1}{N}\langle u_\perp({\bf k})u_\perp({-\bf k})\rangle.
\label{sf.7}
\end{equation}

\begin{figure}[htbp]
\begin{center}
\includegraphics[width=17cm,keepaspectratio,clip=true]{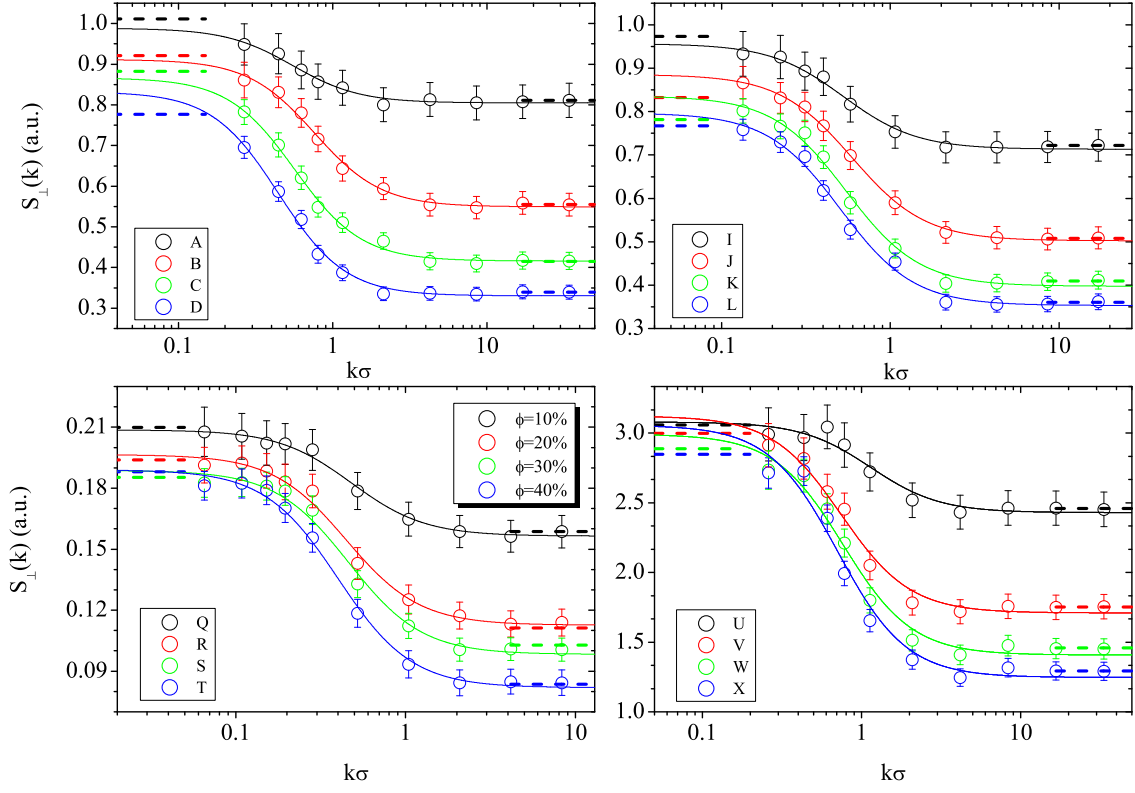}
\end{center}
\caption{Transverse velocity structure factors at various $\phi$ (see
  the legend), with particles' diameters $\sigma=2$ mm (top left),
  $\sigma=1$ mm (top right and bottom left) and $\sigma=4$ mm (bottom
  right), for regions of interest ``center'' (top panels) and
  ``whole'' (bottom panels), and driving with $f=200$ Hz. For
  all experimental parameters, see Table~\ref{t:times}. Dashed lines
  mark the value of $N\langle V_{cm}^2 \rangle$ and $T_g$ at low and
  large values of $k$ respectively. The continuous curves are fits with formula~\eqref{eq:Sklowk}. }
\label{f:sperp_phi}
\end{figure}

\begin{figure}[htbp]
\begin{center}
\includegraphics[width=8cm,keepaspectratio,clip=true,angle=0]{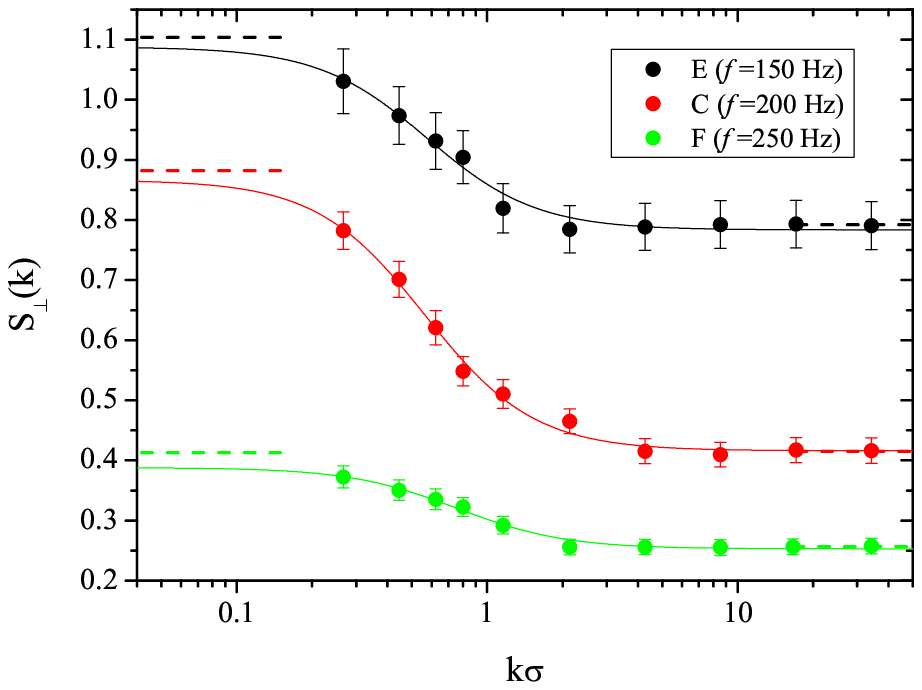}
\includegraphics[width=8cm,keepaspectratio,clip=true,angle=0]{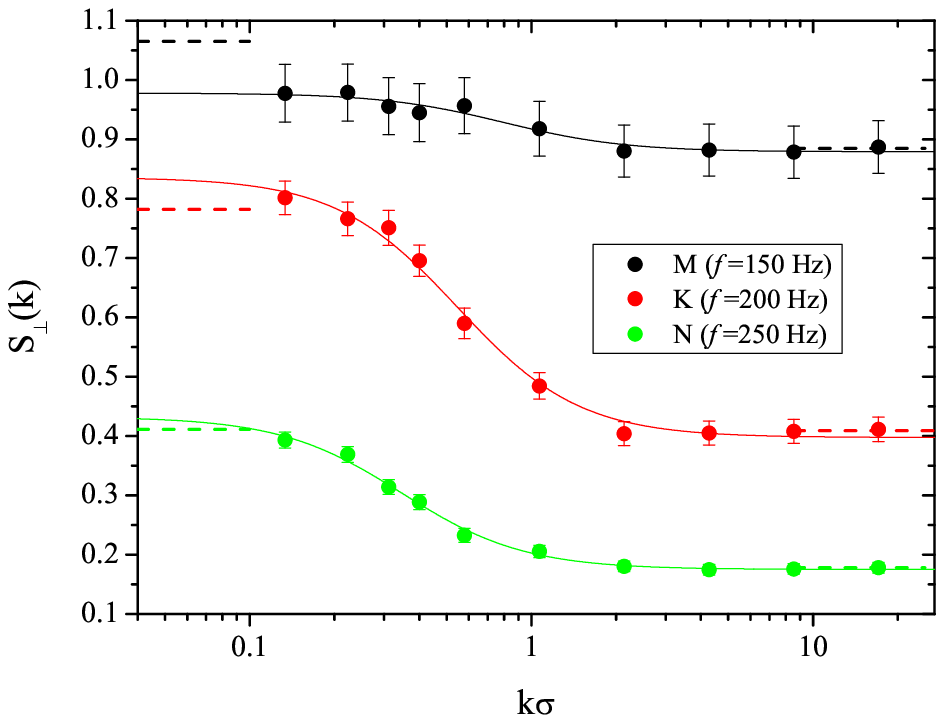}
\end{center}
\caption{Transverse velocity structure factors at various frequencies
  $f$ and fixed packing fraction $\phi=30\%$ and acceleration, with
  particles' diameter $\sigma=2$ mm (left panel) and $\sigma=1$ mm
  (right panel). For all other experimental parameters, see
  Table~\ref{t:times}. Dashed lines mark the value of $N\langle
  V_{cm}^2 \rangle$ and $T_g$ at low and large values of $k$
  respectively. The continuous curves are fits with formula~\eqref{eq:Sklowk}.}
\label{f:sperp_omega}
\end{figure}

\begin{figure}[htbp]
\begin{center}
\includegraphics[width=8cm,keepaspectratio,clip=true,angle=0]{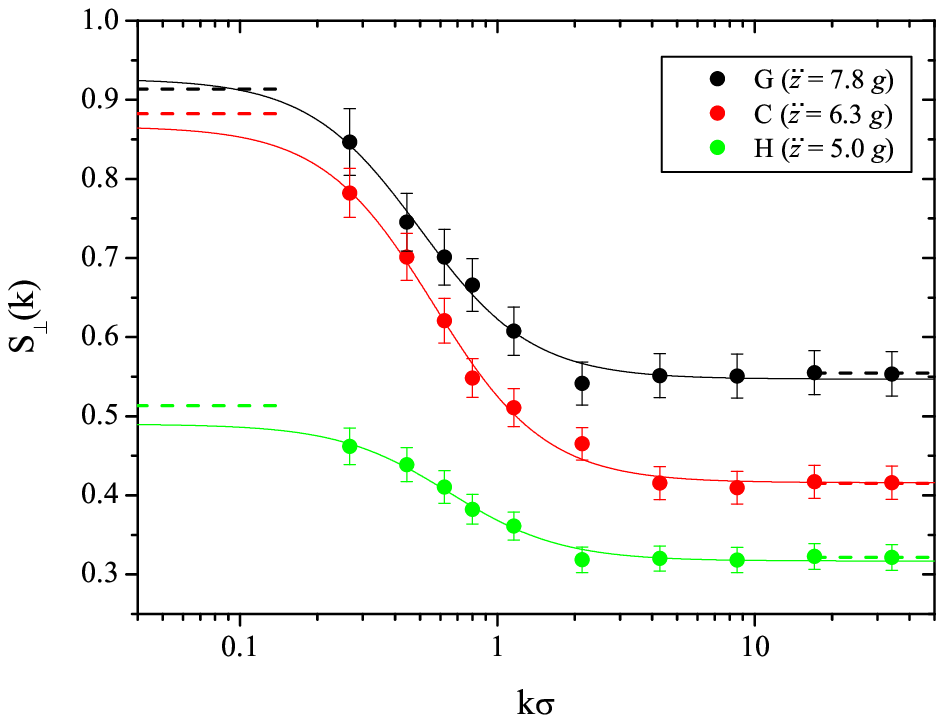}
\includegraphics[width=8cm,keepaspectratio,clip=true,angle=0]{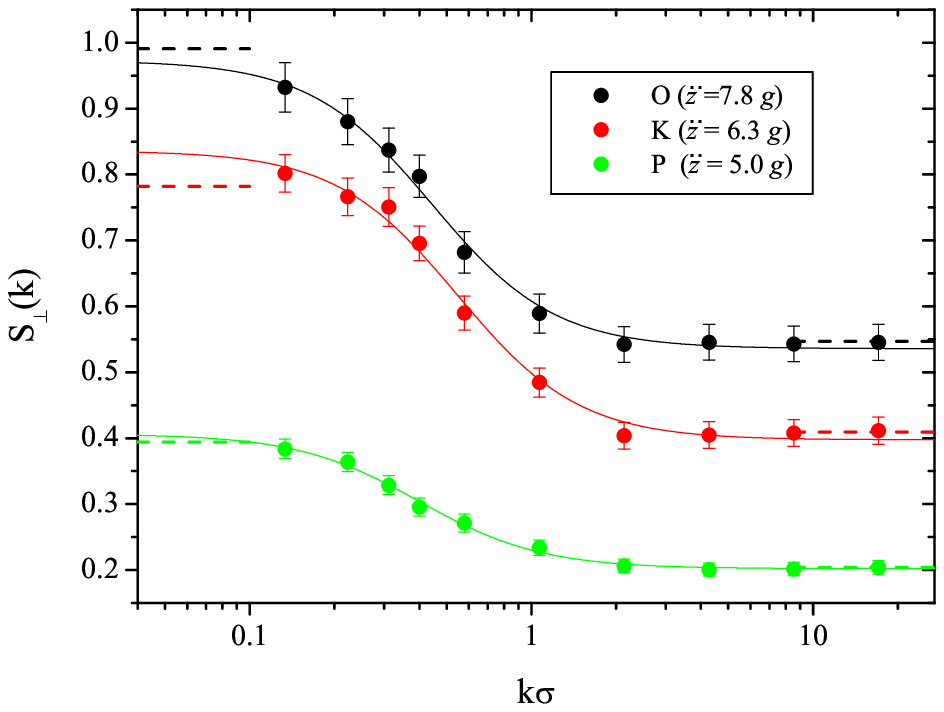}
\end{center}
\caption{Transverse velocity structure factors at various maximum
  acceleration of the plate $\ddot{z}_{max}$ and constant frequency
  $f=200$ Hz, at packing fraction $\phi=30\%$ and with particles'
  diameter $\sigma=2$ mm (left panel) and $\sigma=1$ mm (right
  panel). For all other experimental parameters, see
  Table~\ref{t:times}. Dashed lines mark the value of $N\langle
  V_{cm}^2 \rangle$ and $T_g$ at low and large values of $k$
  respectively. The continuous curves are fits with formula~\eqref{eq:Sklowk}.}
\label{f:sperp_acc}
\end{figure}

In Fig.~\ref{f:sperp_phi}, $S_{\perp}(k)$ is reported for
different values of $\phi$, at fixed amplitude and frequency of
oscillation. In all cases, we find two plateau, at large and small
$k$.  The former coincides with the granular temperature $T_g$ of the
system and is determined by the packing fraction. The latter is much
less dependent on $\phi$ and is approximately constant within
experimental errors. Its value is expected to
be mostly related to the energy injection mechanism, namely to the
shaking parameters.  Indeed, in Figs.~\ref{f:sperp_omega}
and~\ref{f:sperp_acc} the structure factor of transverse modes is
shown upon varying the frequency and amplitude of oscillation,
respectively, keeping fixed the packing fraction. In these cases a
strong dependence of $S_{\perp}(k)$ at small $k$ on these
parameters is observed. At small $k$ we have also reported (see dashed
lines) the values of $N\langle V_{cm}^2\rangle$, whose relevance at large
scales is discussed below.

\begin{figure}[htbp]
\begin{center}
\includegraphics[width=8cm,keepaspectratio,clip=true]{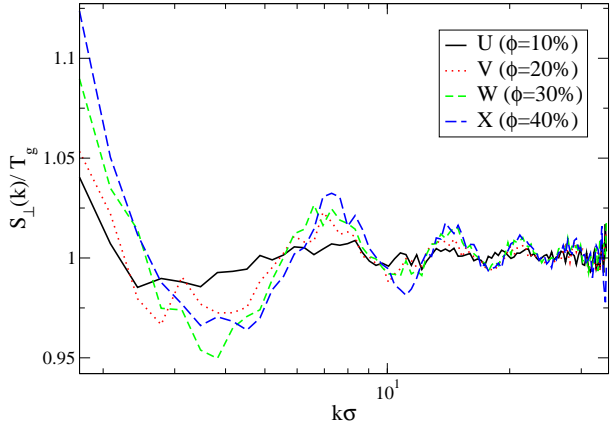}
\end{center}
\caption{Transverse velocity structure factors, rescaled by $T_g$, at
  various packing fractions with particles' diameter $\sigma=4$ mm and
  for driving with $f=200$ Hz (experiments with labels I, J, K, and
  L), at high values of $k$ and finer $k$-resolution, in order to show
  the oscillations, which are similar to the ones observed in the density structure factor. For all
  experimental parameters, see Table~\ref{t:times}. }
\label{f:sperp_fine}
\end{figure}

The main feature arising from the figures illustrated above is a
``sigmoidal'' shape of the structure factor that can be clearly
observed in the whole range of parameters. Note that we have
intentionally avoided a too high resolution in $k$, to neglect
``fine'' effects which have a less clear interpretation: for instance,
it would be interesting but beyond the scope of our analysis here, to
give an interpretation of the oscillations at large k that can be observed with a higher
$k$-resolution, see Fig.~\ref{f:sperp_fine} for an example. The main
``sigmoidal'' shape is remarkably different from what is observed in
elastic fluids, where static velocity structure factors are flat.  The
profile we observe connects two regions in $k$ space with different
energy scales: one associated with the granular temperature and the
other one related to the external driving mechanism.  As it will be
thoroughly discussed below, this behavior suggests the existence of a
\emph{characteristic length-scale} in the system, marking the passage
from one region to the other.
 
\begin{figure}[htbp]
\begin{center}
\includegraphics[width=12cm,keepaspectratio,clip=true]{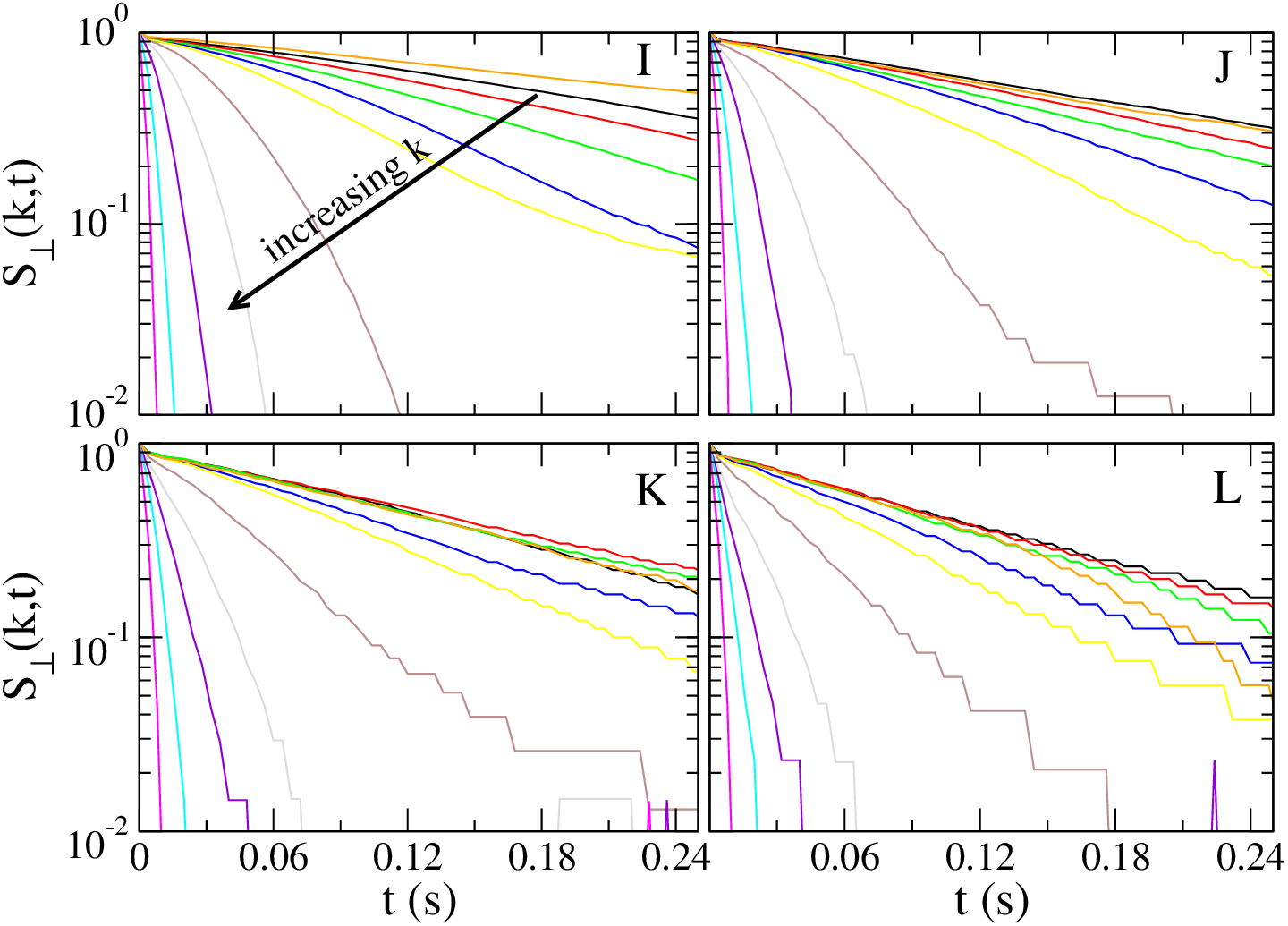}
\end{center}
\caption{Temporal decay of the transverse velocity modes,
  $S_\perp(k,t)$, rescaled by its value at time $t=0$, in experiments
  with different packing fractions $\phi=10\%$ (I), $\phi=20\%$ (J),
  $\phi=30\%$ (K) and $\phi=40\%$ (L), with particles' diameter
  $\sigma=1$ mm and for driving with $f=200$ Hz. For all other
  experimental parameters, see Table~\ref{t:times}. The values of $k
  \sigma$ are the following: $0.089$, $0.14$, $0.2$, $0.26$, $0.32$,
  $0.51$, $1$, $2$, $4$, $8$, $16$. Error bars are omitted to improve legibility.}
\label{f:sperp_ac}
\end{figure}

In order to study the characteristic times in our system, we have also
measured the dynamic structure factor
\begin{equation}
S_{\perp}({\bf k},t)=\frac{1}{N}\langle u_\perp({\bf k},t)u_\perp({-\bf k},0)\rangle, 
\label{sf.8}
\end{equation}
that is reported in Fig.~\ref{f:sperp_ac}, for different values of the
experimental parameters. Notice that, at small $k$, the decay in time
of $S_{\perp}(k,t)$ is nearly exponential. Moreover, the decorrelation
time is greater for smaller $k$. A more detailed analysis of such
behavior is postponed to Section~\ref{temp}.

\begin{figure}[htbp]
\begin{center}
\includegraphics[width=12cm,keepaspectratio,clip=true]{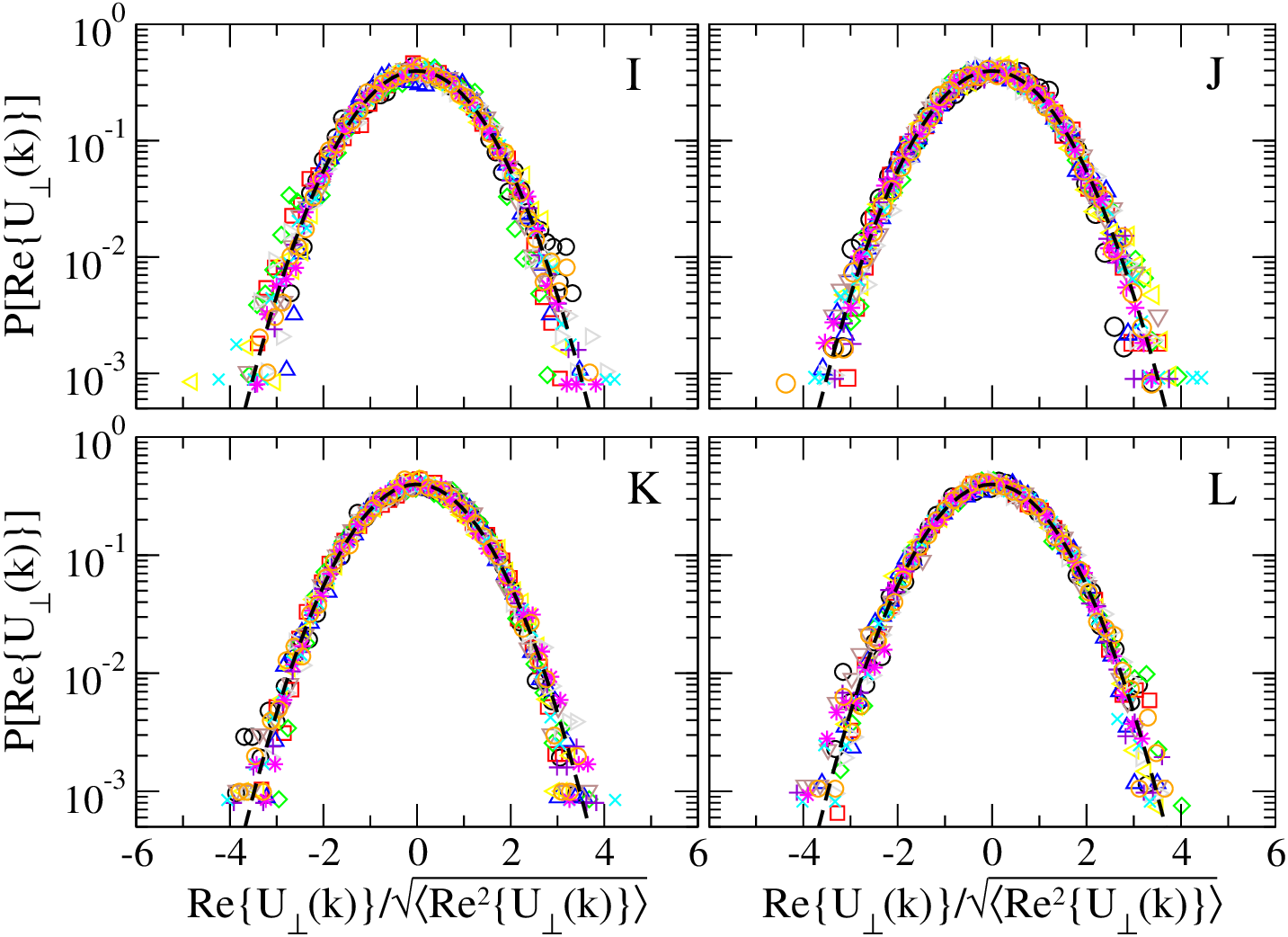}
\end{center}
\caption{Probability density functions of transverse velocity modes
  (real part), rescaled to have unit variance, for experiments with
  different packing fractions $\phi=10\%$ (I), $\phi=20\%$ (J),
  $\phi=30\%$ (K) and $\phi=40\%$ (L), with particles' diameter
  $\sigma=1$ mm and for driving with $f=200$ Hz. For all other
  experimental parameters, see Table~\ref{t:times}. The values of $k
  \sigma$ are the following: $0.089$, $0.14$, $0.2$, $0.26$, $0.32$,
  $0.51$, $1$, $2$, $4$, $8$, $16$.
\label{f:sperp_pdf}}
\end{figure}

In Fig.~\ref{f:sperp_pdf} the probability distribution functions
(rescaled to have unitary variance) of the transverse velocity modes
are shown. For different values of the parameters, we always observe
Gaussian distributions.

%-------------------------------------------------------------------
%                          fig 6
%-------------------------------------------------------------------
%-------------------------------------------------------------------
%-------------------------------------------------------------------

%-------------------------------------------------------------------
%                          fig 7
%-------------------------------------------------------------------
%-------------------------------------------------------------------
%-------------------------------------------------------------------

%-------------------------------------------------------------------
%                          fig 8
%-------------------------------------------------------------------
%-------------------------------------------------------------------
%-------------------------------------------------------------------

%-------------------------------------------------------------------
%                          fig 9
%-------------------------------------------------------------------
%-------------------------------------------------------------------
%-------------------------------------------------------------------

%-------------------------------------------------------------------
%                          fig 10
%-------------------------------------------------------------------
%-------------------------------------------------------------------
%-------------------------------------------------------------------

%-------------------------------------------------------------------
%                          fig 11
%-------------------------------------------------------------------
%-------------------------------------------------------------------
%-------------------------------------------------------------------

%%%%%%%%%%%%%%%%%%%%%
\subsection{Longitudinal velocity modes}

\begin{figure}[htbp]
\begin{center}
\includegraphics[width=12cm,keepaspectratio,clip=true]{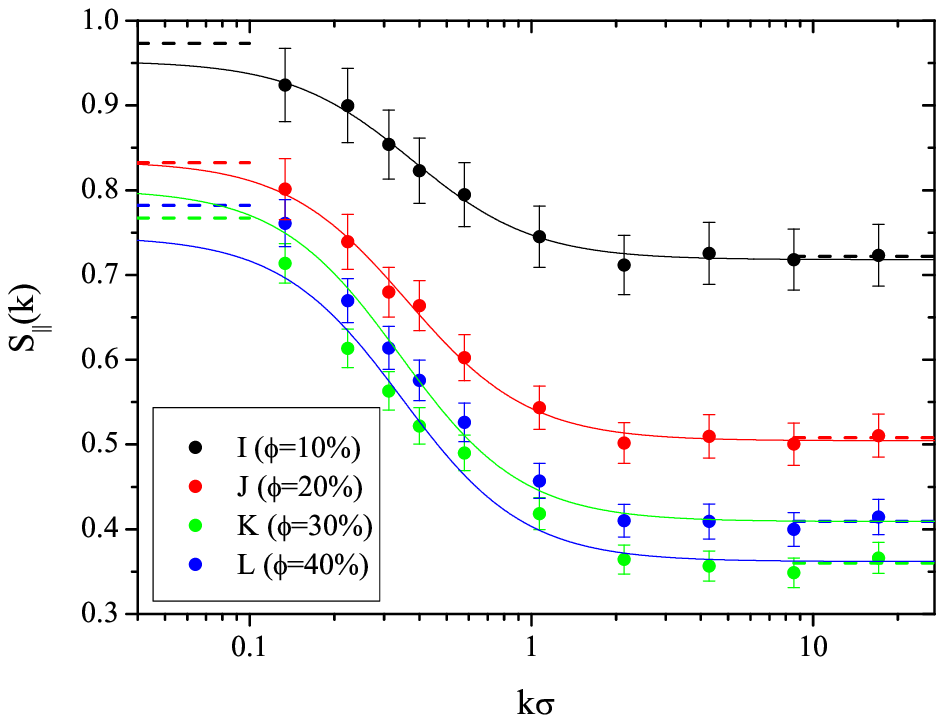}
\end{center}
\caption{Longitudinal velocity structure factors at various $\phi$ (experiments
  with labels I, J, K and L). For all experimental parameters, see
  Table~\ref{t:times}. Dashed lines mark the value of $N\langle
  V_{cm}^2 \rangle$ and $T_g$ at low and large values of $k$
  respectively. The continuous curves are fits with formula~\eqref{eq:Sklong}. }
\label{f:spar_phi}
\end{figure}

The structure factor  of longitudinal velocity modes

\begin{equation}
S_{\parallel}({\bf k})=\frac{1}{N}\langle u_\parallel({\bf k})u_\parallel({-\bf k})\rangle 
%=\frac{1}{Nn^2} \sum_{i=1}^{N} 
%\left\langle \left| ({\bf v}_i\cdot\hat{k})\exp(-i \k \cdot \r_i) \right|^2 \right\rangle,
\label{sf.9}
\end{equation}
is shown in Fig.~\ref{f:spar_phi}, for different values of the
experimental parameters (see caption). The same qualitative behaviour
is observed as for transverse modes: two energy scales are present at
large and small $k$, representing the granular temperature and the
energy scale associated to the fluctuations of the center-of-mass
velocity, respectively.  Again, a characteristic length scale can be
associated with the spatial correlation of longitudinal modes, as
discussed in more detail below.

\begin{figure}[htbp]
\begin{center}
\includegraphics[width=12cm,keepaspectratio,clip=true]{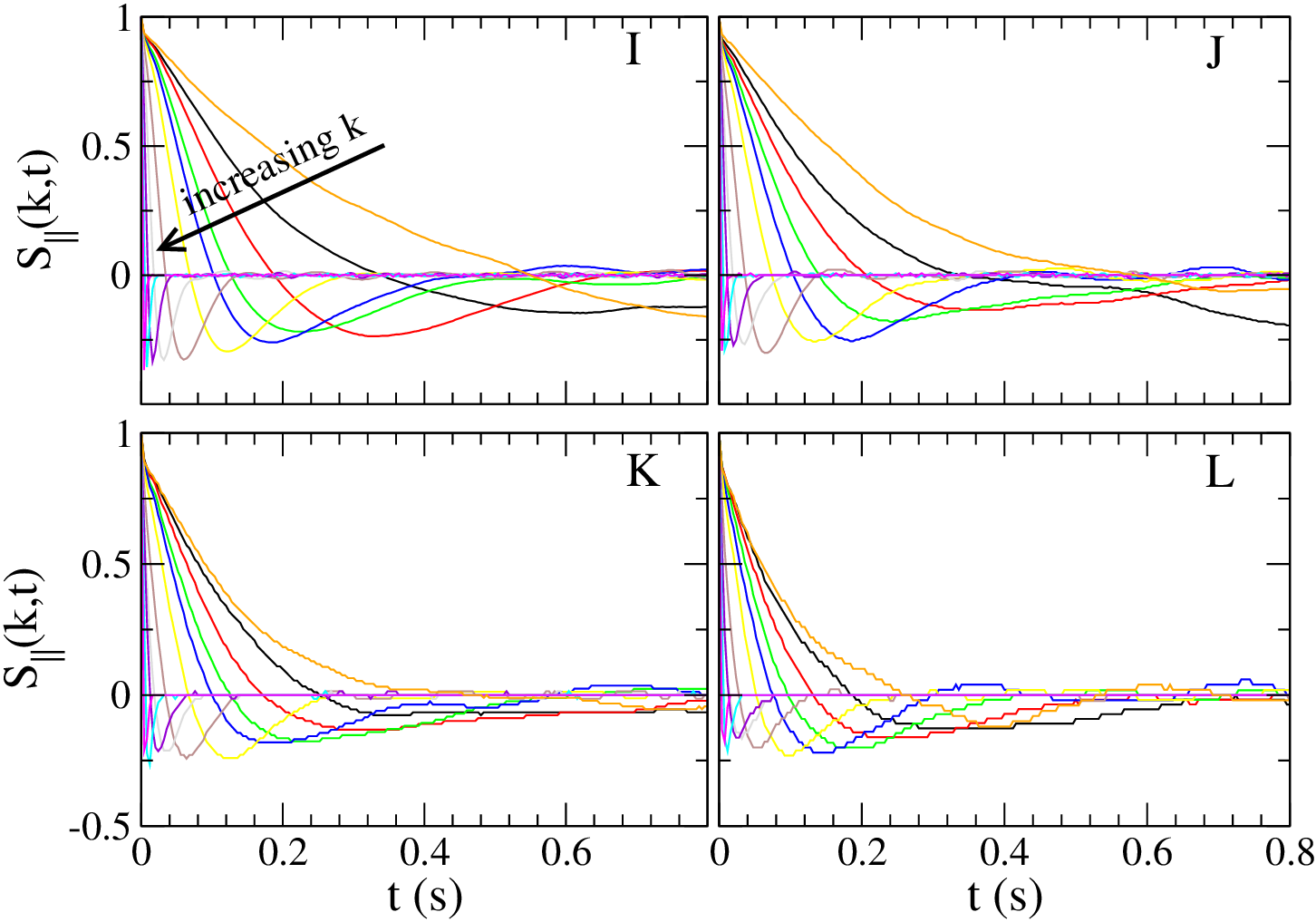}
\end{center}
\caption{Temporal decay of the longitudinal velocity modes,
  $S_\parallel(k,t)$, rescaled by its value at time $t=0$, in
  experiments with different packing fractions $\phi=10\%$ (I),
  $\phi=20\%$ (J), $\phi=30\%$ (K) and $\phi=40\%$ (L), with
  particles' diameter $\sigma=1$ mm and for driving with $f=200$
  Hz. For all other experimental parameters, see
  Table~\ref{t:times}. The values of $k \sigma$ are the following:
  $0.089$, $0.14$, $0.2$, $0.26$, $0.32$, $0.51$, $1$, $2$, $4$, $8$,
  $16$. Error bars are omitted to improve legibility.}
\label{f:spar_ac}
\end{figure}

The temporal decay of time correlations of longitudinal modes
$S_{\parallel}({\bf k},t)=\frac{1}{N}\langle u_\parallel({\bf
  k},t)u_\parallel({-\bf k},0)\rangle$, is shown in
Fig.~\ref{f:spar_ac}. Here we find a behaviour much more complex than
a simple exponential decay. This is due to the coupling of the
longitudinal modes with the other hydrodynamic fields.

\begin{figure}[htbp]
\begin{center}
\includegraphics[width=12cm,keepaspectratio,clip=true]{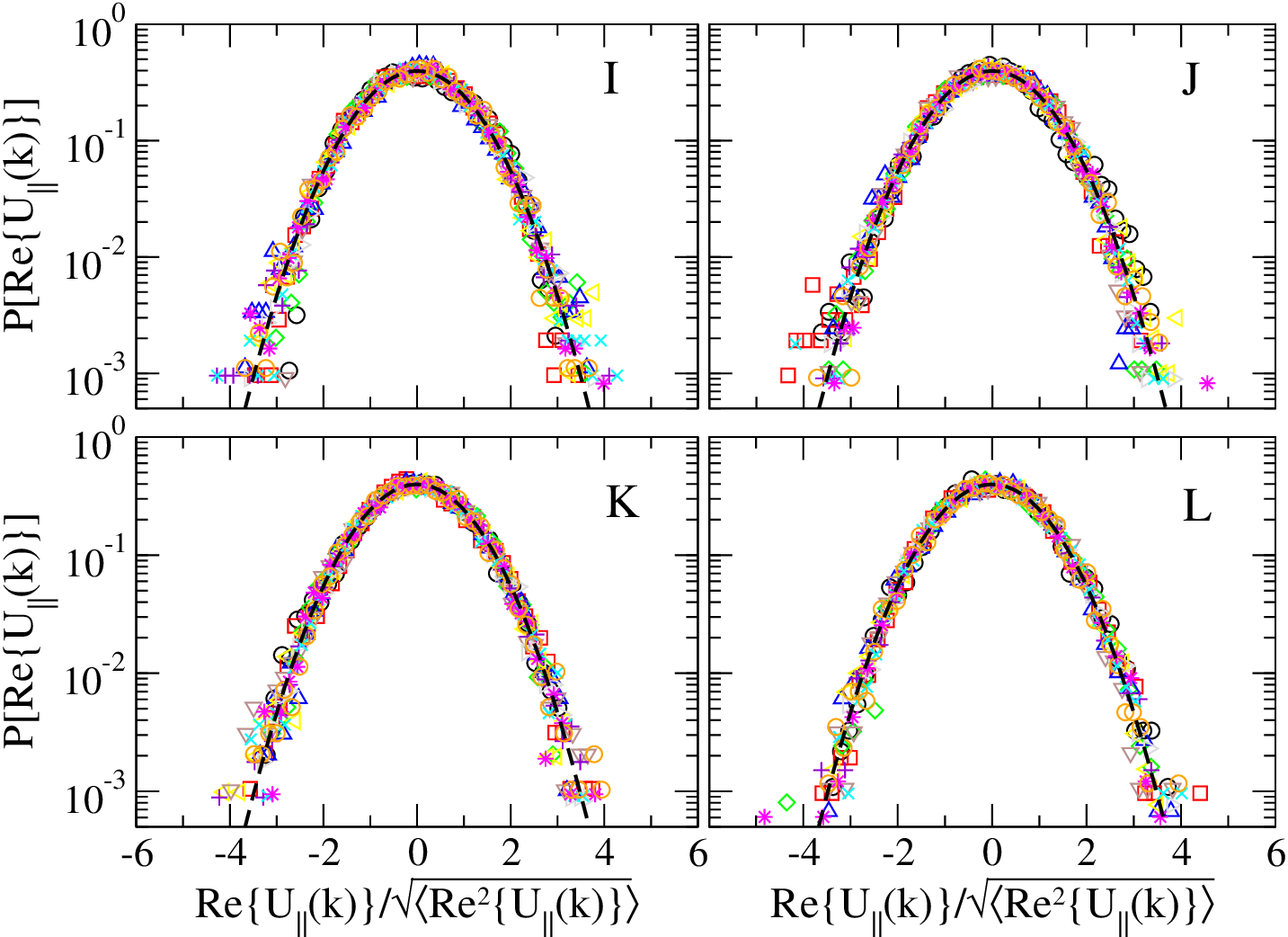}
\end{center}
\caption{Probability density functions of longitudinal velocity modes
  (real part), rescaled to have unit variance, for experiments with
  different packing fractions $\phi=10\%$ (I), $\phi=20\%$ (J),
  $\phi=30\%$ (K) and $\phi=40\%$ (L), with particles' diameter
  $\sigma=1$ mm and for driving with $f=200$ Hz. For all other
  experimental parameters, see Table~\ref{t:times}. The values of $k
  \sigma$ are the following: $0.089$, $0.14$, $0.2$, $0.26$, $0.32$,
  $0.51$, $1$, $2$, $4$, $8$, $16$.
\label{f:spar_pdf}}
\end{figure}

Finally, in Fig.~\ref{f:spar_pdf} the probability distribution functions of
$u_\parallel({\bf k})$ are reported.  Again, we find that such
distributions are nearly Gaussian.

%-------------------------------------------------------------------
%                          fig 12
%-------------------------------------------------------------------
%-------------------------------------------------------------------
%-------------------------------------------------------------------

%-------------------------------------------------------------------
%                          fig 13
%-------------------------------------------------------------------
%-------------------------------------------------------------------
%-------------------------------------------------------------------

%-------------------------------------------------------------------
%                          fig 14
%-------------------------------------------------------------------
%-------------------------------------------------------------------
%-------------------------------------------------------------------

%%%%%%%%%%%%%
\subsection{Comparison with theory and simulations:  ``bath temperature'' and correlation lengths}
\label{compare}

In order to discuss a theoretical model for our system, we show,
starting from the definition of $S_\perp(k)$, see Eq.~(\ref{sf.7}),
what the physical meaning of the two plateau appearing in
Fig.~\ref{f:sperp_phi} is.  At large values of $k$ in Eq.~(\ref{sf.7})
only the diagonal term survives:
\begin{equation}
 S_\perp(k) \rightarrow \langle \frac{1}{N}\sum_{i}\frac{{\bf v}_i^2}{2} \rangle =  T_g,
\label{struct-fact}
\end{equation}
whereas at small values of $k$ we have 
\begin{equation}
S_\perp(k\to 0) = \frac{N}{2} \langle \left( \frac{1}{N}\sum_{i}{\bf v_i} \right)^2 \rangle =
\frac{N}{2} \langle {\bf V}_{cm}^2 \rangle \equiv N \langle V_{cm}^2\rangle.
\end{equation}
For large values of $k$ the velocity structure
factor is proportional to the average square velocity of particles,
i.e. $T_g$, while at small $k$ values one has $S_\perp(k)=N \langle
{\bf V}_{cm}^2 \rangle$, with ${\bf V}_{cm}$ almost independent of
(momentum conserving) collisions and affected only by the interaction
with the vibrating plate. Let us note that, since in our
  system the momentum is not conserved, we have that the value of
  $S_\perp(k=0)$ is not constrained to a $\delta(k)$ and can be
  compared to the values of the function in the neighbourhood of
  $k=0$.

A model theory that fully accounts for the dissipative interactions
among hard particles and between particles and the vibrating vessel is
beyond our scope~\cite{BDKS98,BMG09,MST09}.  We are interested into an effective model
that can reproduce the behavior of collective modes at large length
scales.

For instance, let us consider the shear modes, which, according to
linear hydrodynamic theory, are decoupled from the other modes of the
system \cite{HMD96}.  The presence of an exponential decay for
$\langle u_\perp({\bf k},t) u_\perp(-{\bf k},0) \rangle$, which can be
seen in the panels of Fig.~\ref{f:sperp_ac}, and the measure of a
Gaussian distribution for the real (and imaginary) part of the
variable $u_\perp(k)$, see Fig.~\ref{f:sperp_pdf}, suggest that the
dynamics of $u_\perp(k)$ is ruled by a simple linear Langevin
equation.  Similarly, for the longitudinal modes $u_\parallel(k)$ a
Gaussian distribution for the real (and imaginary) part is found, as
shown in Fig.~\ref{f:spar_pdf}, while the decay in time presents an
exponentially damped oscillation. This behaviour is consistent, as
discussed below, with a linear system of Langevin equations which
couples longitudinal modes to temperature and density modulations, as
dictated by linear hydrodynamics.

One of the central points to be addressed in order to give a
theoretical description of our system is the energy
injection mechanism. Indeed, in a system of particles with dissipative
interactions, the collective behaviour strongly depends on the kind of energy
injection, as it is clear comparing, for instance, the low $k$
tails of velocity structure factors in Refs.\cite{GSVP11,NETP99},
where different kinds of model thermostats are studied.  In
particular, we assume that the velocity of the $i$-th particle
${\bf v}_i$ of our experiment evolves in time according to the stochastic
equation:
\begin{equation}
\dot{{\bf v}}_i(t)=-\gamma_b {\bf v}_i(t) + \boldsymbol{\zeta}_{i}(t)+{\bf F}_i(t),
\label{langevin}
\end{equation}
where the force ${\bf F}_i(t)$ accounts for the inelastic collisions with
other particles and we assume that the interaction between the
vibrating vessel and each particle can be modeled as an equilibrium
thermostat with a viscous drag $-\gamma_b {\bf v}_i(t)$ and a white noise $\boldsymbol{\zeta}_i$
satisfying the 2nd kind FDT, namely we assume
for the noise: $\langle
\boldsymbol{\zeta}_{i}(t) \boldsymbol{\zeta}_{j}(t') \rangle = 2
\gamma_b T_b\delta(t-t')\delta_{i,j}$, with $\gamma_b$ the friction
coefficient.  According to this model, the energy of the center of
mass of the system and the low $k$ value of $S_\perp(k)$ are related
to the temperature of the thermostat:
\begin{equation}
S_\perp(k) \to N \langle V_{cm}^2 \rangle =  T_b. 
\end{equation}

To explain the shape of the structure factors, a theory for
macroscopic fields has to be developed from the microscopic
equation~\eqref{langevin}.  Hydrodynamics is the proper candidate when separation of scales exists between ``fast''
relaxing microscopic modes and ``slow'' relaxating macroscopic ones,
a condition always to be verified in granular fluids~\cite{G99,K99},
but well satisfied in our experiment. Fluctuating hydrodynamics is a
correction to such a theory when macroscopic scales are not large
enough to neglect fluctuations. A more detailed derivation of the
stochastic equations for the hydrodynamic fields starting from
microscopic dynamics can be found in~\cite{GSVP11}. Here, we just
provide phenomenological arguments for the Langevin equation that
governs the shear modes.  According to the observation of two relevant
energy scales in our system, we assume two sources of hydrodynamic
noise of different amplitude: an internal one, $\eta_I(k,t)$, due to
the collisions between particles, and an external one, $\eta_E(k,t)$,
due to the action of the external thermostat. These noises have zero
average and variance:
\begin{eqnarray}
\langle \eta_I(k,t) \eta_I(k',t')\rangle = 2T_g \nu k^2 \delta(t-t') \delta(k+k')\nonumber \\ 
\langle \eta_E(k,t) \eta_E(k',t')\rangle = 2\gamma_b T_b \delta(t-t') \delta(k+k') \nonumber \\
\langle \eta_E(k,t) \eta_I(k',t')\rangle = 0,
\end{eqnarray}  
where $\nu$ is the kinematic viscosity.  Then, in agreement with a
2nd kind FDT principle, we introduce a frictional term proportional
to each source of randomness, obtaining the following Langevin
equation:
\begin{equation}
\dot{u}_\perp(k,t) = - (\gamma_b + \nu k^2 ) u_\perp(k,t) + \eta_I(k,t) + \eta_E(k,t).
\label{eq:langeq}  
\end{equation} 
Such an equation accounts for all the experimental observations on
shear modes presented above, namely the exponential decay of the time
autocorrelation, the Gaussian distribution of
$\textrm{Re}[u_\perp(k,t)]$ (and its imaginary part), and in
particular the shape of the structure factor.  We can say that each
mode is at equilibrium, but with a different temperature.  The system
as a whole is of course out of equilibrium as long as $T_g \neq T_b$.
Such an observation gives us the way to discuss one of the most
intriguing feature of the non-equilibrium structure of velocity
correlations, namely the finding of a correlation length not present
at equilibrium.  Indeed, the temperature spectrum of modes, which is
nothing but the velocity structure factor, can be obtained from
Eq.~(\ref{eq:langeq}) and reads as
\begin{equation}
S_\perp(k)=N^{-1}\langle |u_\perp(k)|^2\rangle = \frac{\gamma_b T_b +
  \nu k^2 T_g}{\gamma_b + \nu k^2 } = T_g + \frac{T_b-T_g}{1+\xi_\perp^2
  k^2},
\label{eq:Sklowk}
\end{equation}
which defines the correlation length for transverse modes
\begin{equation}
\xi_\perp = \sqrt{\nu/\gamma_b}.
\end{equation}
At equilibrium, namely when collisions are elastic and $T_g=T_b$, the
equipartition between modes is perfectly fulfilled and the structure
factor becomes flat, i.e. we have a single temperature for all the
modes: $S_\perp(k)=T_b$.  Differently, in our non-equilibrium system
of interest, where $T_g \ne T_b$, equipartition is broken and we find
an explicit dependence of $S_\perp(k)$ on $k$. Performing the 2D
inverse Fourier transform $\mathcal{T}^{-1}$, in real space
Eq.~(\ref{eq:Sklowk}) acquires a plain meaning:
\begin{equation}
\mathcal{T}^{-1}S_\perp= T_g \delta^{(2)}({\bf r}) + (T_b- T_g)\frac{K_0(r/\xi_\perp)}{\xi_\perp^2},
\label{eq:Sx} 
\end{equation} 
where $K_0(x)$ is the 2nd kind modified Bessel function that, for
large distances, decays exponentially
\begin{equation}
K_0(r/\xi_\perp)\approx \sqrt{\frac{\pi}{2}}\frac{e^{-r/\xi_\perp}}{(r/\xi_\perp)^{1/2}}.
\label{eq:Sx2} 
\end{equation} 
The existence of two different and well defined energy scales has a
neat consequence both in real and in Fourier space. In real space the
distance from equilibrium $T_b-T_g$ determines the amplitude of non
equilibrium correlations in the velocity field, see Eq.~(\ref{eq:Sx}). 
In Fourier space a division in two subsystems
that respectively equilibrated at high or low temperature takes place. By looking
at Fig.~\ref{f:sperp_phi} we may identify the characteristic momentum $k^*$
corresponding to the characteristic length $\xi_\perp$ as the conventional
point which separates the subset of ``hot'' modes from the subset of
``cold'' modes.

For the interested reader, a comprehensive discussion on the relations
between noises, covariances and large scale correlations for the
model~(\ref{langevin}) is addressed in~\cite{GSVP11}.  Clearly, the
linearization of hydrodynamic equations is only possible in a certain
range of densities, and, in particular, when the granular fluid
becomes too dense, new couplings appear and non-linear effects become
relevant. For instance, at packing fractions where a glassy-like
arrest of dynamics occurs~\cite{KAGD07,KSZ10}, the linearized
hydrodynamics is no more expected to be effective.
For the longitudinal velocity modes, we find an exponentially damped
oscillating function. This is consistent with the fact that even
within linear hydrodynamics they are coupled to temperature and
density fluctuations and hence their temporal decay has more than one
characteristic time-scale.
Their frequency power spectrum $S_\parallel(k,\omega)$ has
been calculated analytically for the same energy injection mechanism
of Eq.~(\ref{langevin}) in~\cite{GSVP11}. Relying on the fluctuating
hydrodynamic equations, the static longitudinal velocity correlator
turns out to be well approximated at low $k$ values by the functional
form:
\begin{equation}
S_\parallel(k) = T_b - ( T_b - T_g ) \xi_\parallel^2 k^2 + \mathcal{O}(k^4) =
T_g + \frac{T_b-T_g}{1 + \xi_\parallel^2 k^2} +  \mathcal{O}(k^4),   
\label{eq:Sklong}
\end{equation}
where $\xi_\parallel$ does not depend only on the drag coefficient
$\gamma_b$ of the thermal bath and on the longitudinal viscosity
$\nu_l$ but also on other microscopic parameters, as reported
explicitly in~\cite{GSVP11}.  As can be seen from
Fig.~\ref{f:spar_phi}, the sigmoidal shape of $S_\parallel(k)$
obtained from Eq.~(\ref{eq:Sklong}) is in good agreement with
experiments. The fact that the low momentum approximation of
$S_\parallel(k)$ recovers also the large $k$ asymptote is not
surprising: $S_\parallel(k)$ represents the power spectrum of a
projection of the velocity field and is therefore strictly bounded by
the energy scales dominant al large and small $k$.

%%%%%%%%%%%%%%%%%%%%%%%%%%%%%%%%%%%%%%%%%%%%%%%%%%%%%%%%%%%%%%%%%%%%
\section{Time-scales, temperatures and correlation lengths}
\label{temp}

%-------------------------------------------------------------------
%                          fig 15
%-------------------------------------------------------------------
\begin{figure}[htbp]
\begin{center}
\includegraphics[width=10cm,keepaspectratio,clip=true]{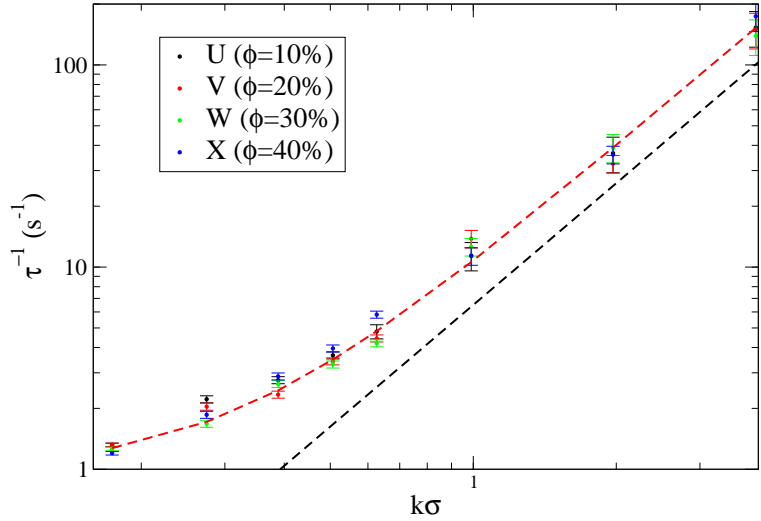}
\end{center}
\caption{Slopes of exponential decays at different values of $k$ for
  $S_\perp(k,t)$ in experiments with different packing fractions
  $\phi=10\%$ (U), $\phi=20\%$ (V), $\phi=30\%$ (W) and $\phi=40\%$
  (X), with particles' diameter $\sigma=4$ mm and for driving with
  $f=200$ Hz. For all other experimental parameters, see
  Table~\ref{t:times}. }
\label{f:slopes}
\end{figure}
%-------------------------------------------------------------------
%-------------------------------------------------------------------

%-------------------------------------------------------------------
%                          fig 16
%-------------------------------------------------------------------
\begin{figure}[htbp]
\begin{center}
\includegraphics[width=10cm,keepaspectratio,clip=true]{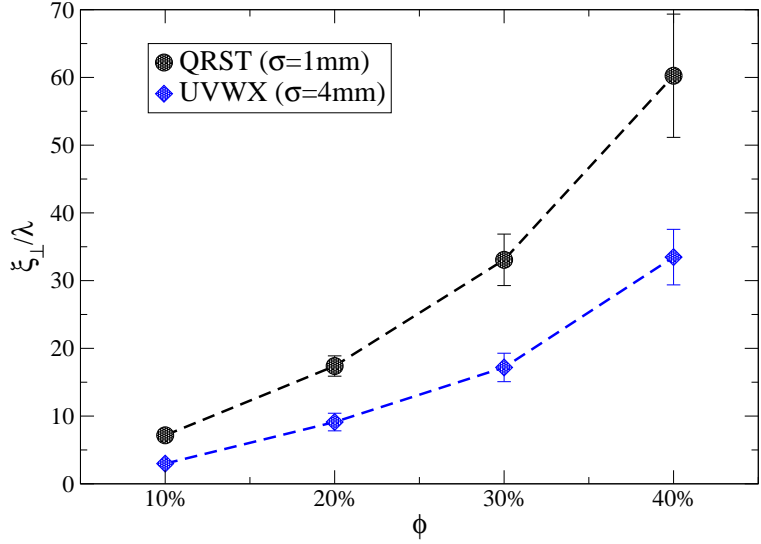}
\end{center}
\caption{Correlation lengths $\xi_\perp$ rescaled by the mean free
  path $\lambda$ as function of the packing fraction $\phi$, for the
  transverse velocity fields for different groups of experiments, with
  labels Q, R, S and T (diameter $\sigma=1$ mm) and with labels U, V,
  W and X (diameter $\sigma=4$ mm), for driving with $f=200$ Hz. The
  method of computing $\xi_\perp$ is discussed in Sec.~\ref{temp}. For
  all other experimental parameters, see
  Table~\ref{t:times}. }
\label{f:lung}
\end{figure}
%-------------------------------------------------------------------
%-------------------------------------------------------------------

From Eq.~(\ref{eq:langeq}) the theoretical prediction for the time
autocorrelation is 
\begin{align}
S_\perp(k,t)&\sim \exp(-t/\tau(k))\\
\tau(k)^{-1}&= \gamma_b + \nu k^2. \label{tau}
\end{align}
The decay in time of $S_\perp(k,t)$, as reported for instance in
Fig.~\ref{f:sperp_ac}, can be fit by exponentials, accepting a
relative error in the order of $5\div 10\%$ after a least squares
procedure. For a few experiments we have analyzed the results of such
fits, finding good agreement with the parabolic behaviour, see
Fig.~\ref{f:slopes}.  From the parabolic fit we extract both the
kinematic viscosity $\nu$ and the drag coefficient of the vibrating
vessel $\gamma_b$ and obtain the length-scale $\xi_\perp$ at various packing
fraction. Those values, for some of the experiments, are reported in
Table~\ref{t:decays}.

\begin{table}
\begin{tabular}{|c||c|c||c|}
 \hline  $\phi$  & $\gamma_b$(s$^{-1}$) & $\nu_\perp$(mm$^2$/s) & $\xi_\perp$(mm) \\ \hline
\hline  10\%  &$1.1\pm 0.1$  &$170 \pm 20$   &$11.8 \pm 2$ \\ \hline
  20\%  &$1.1\pm 0.1$  &$145 \pm 20$  &$11.5 \pm 2$ \\ \hline
  30\%  &$0.97\pm 0.1$ &$156 \pm 20$ &$12.5 \pm 2$ \\ \hline
  40\%  &$0.96\pm 0.1$ &$145 \pm 20$ &$12.3 \pm 2$\\ \hline
\end{tabular}
\caption{$\gamma_b$, $\nu_\perp$ and $\xi_\perp$ as measured from
  exponential decays, formula~\eqref{tau}, in experiments with
  particles' diameter $\sigma=4$ mm, for driving with $f=200$ Hz and
  all the other parameters are fixed to the values corresponding to
  label U of Table~\ref{t:times}. In particular, the dependence of
  $\xi_\perp/\lambda$ on $\phi$ is shown in
  Fig.~\ref{f:lung}. \label{t:decays}}
\end{table}

Both $\gamma_b$ and $\nu_\perp$ remain roughly constant when the
packing fraction is varied from $10\%$ to $40\%$.  A constant value
for $\gamma_b$ is in fair agreement with the interpretation that it
represents an effective viscous drag of the particle-plate
interaction, as it appears in the microscopic ``thermostatted'' model,
Eq.~\eqref{langevin}. The kinematic viscosity, on the other side,
could weakly depend on $\phi$, as well as on the internal temperature
$T_g$, but variations of those parameters are not large enough to
result in appreciable variations of $\nu_\perp$. Estimates, expected
to be reliable in the dilute limit, give $\nu_\perp \sim T_g \tau_c$
where $\tau_c$ is the mean free time between collisions: the
comparison with the values of $T_g$ (reported below) is compatible
with a mean free time $\tau_c\sim 0.1 \div 0.5$ seconds, which, in
turn, agrees with the typical crossover time from ballistic to
diffusive behavior, see Fig.~\ref{f:diff}. It is also important to
notice that the order of magnitude of time-scale separation between
the single particle relaxation $\gamma_b$ and the collective
(collisional) relaxation $\tau_c$ is $\tau_c\gamma_b \sim 0.1 \div
0.5$: this is reflected in a not very large difference between $T_b$
(or its estimate $N\langle V_{cm}^2\rangle$) and $T_g$.  
  Such lack of good separation of scales makes difficult to use
  linearized hydrodynamics for {\em quantitative} predictions of
  velocity correlations: its predictive power is fully appreciated
  only when time-scales are well separated, as it happens with
  numerical simulations~\cite{GSVP11,GSVP11b}. Here, the fits of
  structure factor data realized with formula~\eqref{eq:Sklowk} and
  shown in
  Figures~\ref{f:sperp_phi},~\ref{f:sperp_omega},~\ref{f:sperp_acc}
  and~\ref{f:spar_phi}, lead to estimates of the physical parameters
  not always consistent with those obtained independently (e.g. in the
  procedure discussed above for $\gamma_b$ and $\nu_\perp$). For this
  reason in the following we have used more rough estimates of the
  main observables when comparing different experiments, as it is
  done in Table ~\ref{t:measures}. 

\begin{table}
\begin{tabular}{|c|c||c|c||c|c|c|c|}
 \hline  Label  & $\phi$ & $T_g$ &  $N\langle V_{cm}^2\rangle \approx T_b$  & $\xi_\perp$  & $\xi_\parallel$  & $\xi_\perp/\lambda$  & $\xi_\parallel/\lambda$ \\[-6pt]
                & & (mm$^2$/s$^2$) & (mm$^2$/s$^2$) & mm & mm & &  

\\ \hline
 \hline   A &10\%  &730  &910  & 19  & 21  &  5  &  5 \\ \hline
  B &20\% &499  &829  & 15  & 20  & 10  & 13 \\ \hline
  C &30\% &373  &794  & 19  & 21  & 22  & 25 \\ \hline
  D &40\% &306  &699  & 20  & 20  & 42  & 41 \\ \hline
  E &30\% &713  &993  & 17  & 19  & 20  & 23 \\ \hline
  F &30\% &231  &371  & 14  & 18  & 17  & 22 \\ \hline
  G &30\% &499  &822  & 20  & 19  & 24  & 23 \\ \hline
  H &30\% &289  &462  & 17  & 18  & 20  & 22 \\ \hline
\hline  I &10\%  &650  &876  & 11  & 16  &  6  &  8 \\ \hline
  J &20\% &457  &749  & 10  & 15  & 13  & 19 \\ \hline
  K &30\% &368  &704  & 11  & 17  & 27  & 40 \\ \hline
  L &40\% &324  &691  & 12  & 17  & 50  & 71 \\ \hline
  M &30\% &796  &959  &  7  & 16  & 17  & 39 \\ \hline
  N &30\% &160  &370  & 15  & 17  & 37  & 42 \\ \hline
  O &30\% &492  &892  & 13  & 17  & 31  & 40 \\ \hline
  P &30\% &183  &355  & 15  & 14  & 36  & 35 \\ \hline
\hline  Q &10\% &600  &794  & 13  & 19  &  7  & 10 \\ \hline
  R &20\% &421  &733  & 14  & 12  & 17  & 16 \\ \hline
  S &30\% &389  &701  & 14  & 21  & 33  & 50 \\ \hline
  T &40\% &316  &711  & 15  &  5  & 60  & 23 \\ \hline
\hline  U &10\% &536  &666  & 22  & 32  &  3  &  4 \\ \hline
  V &20\% &381  &653  & 28  & 33  &  9  & 11 \\ \hline
  W &30\% &318  &629  & 29  & 34  & 17  & 20 \\ \hline
  X &40\% &282  &620  & 32  & 32  & 33  & 33 \\ \hline
\end{tabular}
\caption{\label{t:measures} Temperatures and correlation lengths measured in the experiments, following the phenomenological interpretation given in Section~\ref{compare}. Lengths are measured in millimeters, time is measured in seconds.}
\end{table}

The values for the correlation lengths are
simply obtained by interpolating the wavelength where the structure
factors cross half of their way between their maximum and minimum
value, i.e. $\xi_\perp = 2\pi/k^*$ and $S_\perp(k^*) \approx
\frac{S_\perp(k \to 0)+S_\perp(k \to \infty)}{2}$, and analogously for
$\xi_\parallel$. Such a definition does not coincide in absolute value
with that, coming from the theoretical model, $\xi_\perp \sim
\sqrt{\nu/\gamma_b}$ : indeed the latter is smaller of about a factor $3$
(compare Table~\ref{t:decays} with Table~\ref{t:measures}, last
four lines). Nevertheless, when comparing the same protocol of
measurement in different experimental conditions, we expect the choice of
protocol to be not too relevant.

The granular temperatures $T_g$ are compatible with thermal velocities
$\sqrt{T_g}$ which are about $1/5 \div 1/10$ of the maximum
vertical velocity of the plate, independently of the mass of the
particles. Fluctuations of center of mass velocity, $N\langle
V_{cm}^2\rangle$, remain more stable than $T_g$ when the packing
fraction is changed, while they do show clear variations when the shaking
parameters are modified: this behavior is in good agreement with the
interpretation of such a quantity as the estimate of the ``bath
temperature'' $T_b$. Apparently, changing the vibration frequency (at
constant acceleration) has a more deep effect on $T_b$ with respect to
a variation of the acceleration at constant frequency (compare
experiments E-F vs G-H, using C as a reference, as well as experiments
M-N vs O-P, using K as a reference).

A general comment concerns the fact that the best agreement with the
phenomenological fluctuating hydrodynamic model is achieved for the
experiments with the fully photographed system (i.e. without excluding
the border regions), experiments from $Q$ to $X$. In those cases,
indeed, one can reach smaller values of $k$ both in comparison with the
mean free path $\lambda$ and with the correlation lengths. Moreover, in this case the estimate of
$T_b$ through $N\langle V_{cm}^2\rangle$ is more reliable because the system
under observation is really {\em closed}, i.e. center of mass
fluctuations are not influenced by collisions of particles inside the
system with particles outside the system.

The correlation lengths separating the ``cold'' (inelastic) scales
from the ``hot'' (elastic) ones, do not change appreciably as the
packing fraction is changed and all other parameters are fixed, but
they display a pronounced {\em increase} with the packing fraction
when normalized by the mean free path, as can be clearly seen in
Fig.~\ref{f:lung}. The meaning of such an observation is that the
number of collisions between correlated particles increases with the
occupied volume. The longitudinal correlation length $\xi_\parallel$
is -- on average -- slightly larger than the transveral one
$\xi_\perp$, as expected also from a more detailed analysis of
fluctuating hydrodynamics~\cite{GSVP11}.

%%%%%%%%%%%%%%%%%%%%%%%%%%%%%%%%%%%%%%%%%%%%%%%%%%%%%%%%%%%%%%%%%%%%%%%%

%%%%%%%%%%%%%%%%%%%%%%%%%%%%%%%%%%%%%%%%%%%%%%%%%%%%%%%%%%%%%%%%%%%%%%%%%%%%%
\section{Conclusions}
\label{s:concl}

We have reported experiments on the measurements of density and
velocity structure factors in a homogeneously driven quasi-2D granular
material. We compared the results with a recent fluctuating
hydrodynamic theory derived from a thermostatted microscopic
model~\cite{GSVP11}. To the best of our knowledge, together with
a previous publication which included some of the results presented
here~\cite{GSVP11b}, this is the first systematic comparison between a
granular experiment and granular fluctuating
hydrodynamics. Considering the approximations involved in the theory,
it is remarkable to observe agreements in so many aspects: 1) the basic
shape of the velocity structure factors, which interpolates between a
``high energy'' plateau at small $k$ and a ``low energy'' plateau at
large $k$; 2) the decay of autocorrelation of velocity modes, which is
roughly exponential for transverse modes and damped oscillating for
the longitudinal ones; 3) the Gaussian character of all fluctuations of
modes; 4) the closeness of the high energy plateau to the variance of the
fluctuations of the center of mass velocity; 5) the fact that such a
plateau strongly depends on the shaking parameters and much more
weakly on the packing fraction.

There are two basic assumptions underlying the theory, which are
indirectly confirmed by the good agreement between the experiment and
the predictions of the theory: 1) a microscopic model which includes a
``thermostat'' with temperature $T_b$ and viscous drag $\gamma_b$; 2)
a sufficient separation of scales such that external noise (from the
thermostat) and internal one (from collisions) are separated and the
latter is well reproduced by a ``local equilibrium''
assumption~\cite{OS06}. The ``bath temperature'' $T_b$ , indeed, seems
to be well measured in the experiment by evaluating the high energy
plateau of the velocity structure factors (at low $k$) or the variance
of the fluctuations of the center of mass velocity multiplied by
$N$. However, the separation of scales is not significant enough to
get a perfect comparison for the complete structure factors (e.g. with
formula~\eqref{eq:Sklowk} or~\eqref{eq:Sklong}). The theory allows us
to define and measure a non-equilibrium correlation length as well,
which increases (in terms of particle mean free paths) when the
packing fraction is increased, representing the separation of ordered
``cold'' scales and more disordered ``hot'' scales which are in
equilibrium with the bath.

Future investigations including a more detailed study of the
vertical-to-horizontal energy transfer mechanism are in progress, with the aim of achieving
a predictive theory for the properties of the effective bath $T_b$ and
$\gamma_b$ as functions of the shaking parameters and the frictional
properties of the surface.

%%%%%%%%%%%%%%%%%%%%%%%%%%%%%%%%%%%%%%%%%%%%%%%%%
\begin{acknowledgments} 
A warm acknowledgment is reserved to M D Deen Islaam for helping with
all mechanical problems in the experimental setup.  We also thank
Umberto Marini Bettolo Marconi for fruitful discussions and Teun
Vissers for a careful reading of the manuscript. The work is supported
by the ``Granular-Chaos'' project, funded by the Italian MIUR under
the FIRB-IDEAS grant number RBID08Z9JE.
\end{acknowledgments}

%%%%%%%%%%%%%%%%%%%%%%%%%%%%%%%%%%%%%%%%%%%%%%%%%%%%%%%%%%%%%%%%%%%%%%%%%%%%%
%\section{Appendix A}
%%%%%%%%%%%%%%%%%%%%%%%%%%%%%%%%%%%%%%%%%%%%%%%%%%%%%%%%%%%%%%%%%%%%%%%%%%%%%

%%%%%%%%%%%%%%%%%%%%%%%%%%%%%%%%%%%%%%%%%%%%%%%%%%%%%%%%%%%%%%%%%%%%%%%
%%%%%%%%%%%%%%%%%%%%%%%%%%%%%%%%%%%%%%%%%%%%%%%%%%%%%%%%%%%%%%%%%%%%%%%
%\bibliography{fluct.bib}
%\bibliography{puglisi.bbl}

\end{document}